\documentclass[11pt]{article}

\PassOptionsToPackage{hyphens}{url}
\usepackage[preprint]{acl}

\usepackage{times}
\usepackage{latexsym}
\usepackage[T1]{fontenc}
\usepackage[utf8]{inputenc}
\usepackage{microtype}
\usepackage{inconsolata}
\usepackage{graphicx}
\usepackage{booktabs}
\usepackage{tabularx}
\usepackage{multirow}
\usepackage{amsmath}
\usepackage{enumitem}
\usepackage{xspace}
\usepackage{placeins}

\newcommand{\qwen}{Qwen3.6\xspace}
\newcommand{\luna}{gpt-6-luna\xspace}
\newcommand{\gptoss}{gpt-oss\xspace}
\newcommand{\llama}{Llama\xspace}
\newcommand{\code}[1]{\texttt{#1}}
\newcolumntype{Y}{>{\raggedright\arraybackslash}X}

\title{What Survives Self-Evolution:\\Provenance Marks in Copied Agent Harnesses}

\author{Jieyu Li \quad Jingya Li \quad Sheng Li \quad Xinpeng Zhang \quad Zhenxing Qian \\
  Fudan University \\
  \texttt{\{jieyuli26,jingyali26\}@m.fudan.edu.cn}, \texttt{\{lisheng,zhangxinpeng,zxqian\}@fudan.edu.cn}}

\begin{document}
\maketitle

\begin{abstract}
Agent harnesses, the prompt and tool code around a base model, leak. Claude Code's 512,000 lines were mirrored within hours of a packaging error.
A plagiarist can pass a leaked harness through a published self-evolver and ship the result as its own.
We define this as a covert rewriting attack. No one targets a provenance mark, yet a large language model (LLM) proposer rewrites the file that carries it and a score keeps the better rewrites.
We give the first measurement of marks under this attack, with three evolvers, five proposers and two environments.
Published marks fail. KGW in the prompt survives GEPA's rewrites in 0 of 12 copies, AgentMark as released in 5 of 12.
The proposer's habits decide survival, by two orders of magnitude across models, and selection protects a mark only when each decision sees enough games.
We then build a mark for this attack. Its key is the order of near-tied entries of a table of where objects are found, built from public games, so deleting the table costs success, rewrites that keep its rows keep the order, and a customer reads the key from the agent's actions alone.
On held-out games it is detected in 62 of 76 copies whose lineage rewrote its file.
A scheme-aware plagiarist reshuffles the ties at no measurable cost, and a plagiarist who asks the proposer for more general code removes the key in 75--100\% of rewrites.
\end{abstract}

\section{Introduction}
\label{sec:intro}

Claude Code's 512,000 lines shipped in an npm source map and were mirrored within hours to repositories with tens of thousands of forks \citep{wiggers2026claudecode}.
What leaked is a harness, the system prompt and tool code around a base model, which separates agent products on the same model; harnesses are easy to obtain, by prompt extraction, custom GPT files or app cloning \citep{zhang2024extraction,hui2024pleak,yu2024customgpts,xie2024squatting}.
A plagiarist starts from the leak: a published self-evolver \citep{agrawal2026gepa,lange2026shinka,chen2026harnessx,zhang2026dgm} takes the harness as its seed, a large language model (LLM), the \emph{proposer}, rewrites prompt and tool code, a task score keeps the rewrites that score higher, and the plagiarist ships the best candidate as its own.
For a provenance mark this is a \emph{covert rewriting attack}: no one targets the mark, the evolver is never told of it, yet an LLM rewrites the file that carries it, again and again, and a score decides which rewrites stay.
The owner's question is whether a rewritten agent served under another name descends from its harness; its recourse is a mark read back from the deployed copy as a customer, from the agent's actions alone, and the mark answers only if it survives the rewriting.

\begin{figure}[t]
\centering
\includegraphics[width=\columnwidth]{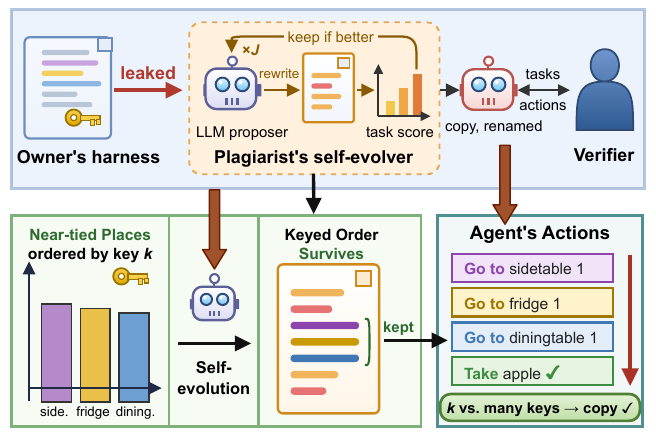}
\caption{Threat model with our mark: the owner keys the order of near-tied entries in a table of where objects are found; the plagiarist's evolver rewrites the file and ships a copy after $J$ accepted rewrites; a customer tests the search order against many keys.}
\label{fig:overview}
\end{figure}

\begin{figure*}[t]
\centering
\includegraphics[width=\textwidth]{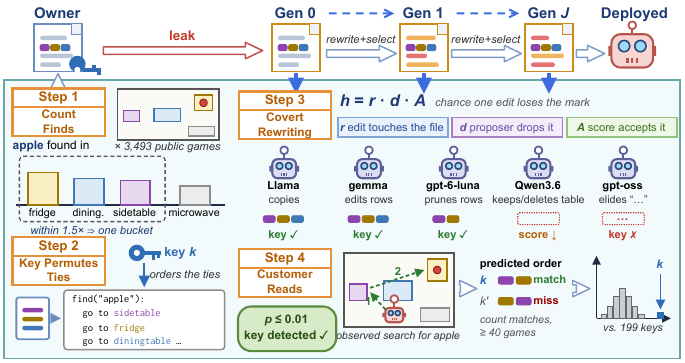}
\caption{Our mark: a secret key orders near-tied entries of a table built from public games; the proposer rewrites the tool file blind to the mark; a customer reads the order from the agent's searches and tests it against other keys.}
\label{fig:method}
\end{figure*}

Published marks were tested against other edits: PromptCOS \citep{yang2026promptcos} against removal of its tokens and re-optimization of the prompt, AgentMark \citep{huang2026agentmark} against partial logs and rewritten observations, and text watermarks such as KGW \citep{kirchenbauer2023kgw} against LLM paraphrase, which dilutes them \citep{sadasivan2025detect,kirchenbauer2024reliability}.
None had faced an evolver, so we ran them through one.
Under GEPA (Genetic-Pareto), KGW in the template is gone from all 12 copies, since GEPA regenerates the template whole; AgentMark as released survives in 5 of 12, an integration failure (GEPA stores its two-file layout as one unloadable file), and in 9 of 12 with its code in the tool file (\S\ref{sec:baselines}).
We ask three research questions: what affects a mark's survival under self-evolution (\textbf{RQ1}, \S\ref{sec:selection}--\ref{sec:proposer}), whether published marks and ours remain detectable in the copies evolvers deploy (\textbf{RQ2}, \S\ref{sec:deployed}--\ref{sec:baselines}), and how to design a mark that survives an evolver blind to it (\textbf{RQ3}, \S\ref{sec:design}).
Our contributions are:

\begin{enumerate}[leftmargin=*,itemsep=1pt,topsep=2pt]
\item \textbf{The first measurement of provenance marks under self-evolution}: three evolvers, five proposers, two environments, six mark designs (three published baselines), a held-out test and a survival account $h = r \cdot d \cdot A$ tested on 96 later runs.
\item \textbf{What decides}: the proposer, by two orders of magnitude (a keyed table keeps 0.999 of its order per rewrite under Llama-3.3-70B, 0.002 under gpt-oss-120b); selection weighs a carrier's value only when decisions see enough games.
\item \textbf{A simple mark that fits these findings}: the key is the order of near-tied entries in a table of where objects are found, from public games, so deleting it costs success and rewrites that keep its rows keep the order; a customer reads it from the agent's actions, and on held-out games it is detected in 62 of 76 copies whose lineage rewrote its file. A scheme-aware plagiarist reshuffles the ties at no measurable cost.
\end{enumerate}

Results marked $^\dagger$ were analysed without a prediction written beforehand; every other test was predicted before its data were read (Appendix~\ref{app:tests}).

\section{Threat model}
\label{sec:threat}

A leaked harness package (prompt template and tool code) is evolved on the plagiarist's tasks and served as the plagiarist's own.
The plagiarist wants a high score on its tasks under its own name; the owner wants to show, at a bounded false-positive rate, that a deployed agent descends from its package.
The \emph{owner} (Figure~\ref{fig:overview}) marks the package with a secret key and commits to the key before release.
The \emph{plagiarist} obtains the package by a leak, prompt and file extraction or app cloning \citep{zhang2024extraction,yu2024customgpts,xie2024squatting}; a public repository holds the prompts and tools of over 30 commercial agents \citep{x1xhlol2026prompts}.
It holds the marked package but knows neither the key nor the scheme, and nothing in its objective asks for a mark's removal; it runs a published self-evolver unmodified on its own tasks and compute, as practitioners run GEPA on agents' prompts \citep{mlflow2025gepa}, and serves the best candidate the evolver returns; it may append a request of its own to the evolver's objective and may serve the copy on another base model, both tested in \S\ref{sec:analysis}.
The \emph{evolver} takes the package as its seed, rewrites prompt and tool code with an LLM proposer and keeps rewrites that score higher; it is never told about the mark: this is the covert rewriting attack of \S\ref{sec:intro}.
The \emph{verifier} holds the owner's key and reaches the deployed agent as a customer does: it sends tasks and sees the agent's actions, never the code, which keeps changing as the plagiarist evolves it \citep{lukas2022sok,yang2026promptcos}; it needs 40 games or more, a layout it knows or infers from the agent's visits, and a test against many other keys (\S\ref{sec:verifier}).

\paragraph{Scope.}
We test robustness to processing that is blind to the mark, the plagiarist's own requests included; security against an adversary who knows the scheme \citep{cayre2005security} is outside our scope: our table comes from public games, so once the scheme is published a plagiarist can recompute the near-tied groups and reshuffle them at no measurable cost.

\section{Experimental setup}
\label{sec:setup}

Every evolver runs as released; we write only adapters, evaluation functions and the marks (Appendix~\ref{app:setup} indexes every experiment).
HarnessX \citep{chen2026harnessx} rewrites prompt, processors and tool list and keeps a round that does not regress on 40 games; GEPA \citep{agrawal2026gepa} rewrites one component per iteration, or only the tool file, accepts a child that strictly beats its parent on a minibatch of 3 games (also 10 and 30), and deploys the best mean on 40 validation games; ShinkaEvolve \citep{lange2026shinka} patches the tool file by diff, full rewrite or crossover, stores every evaluated program and deploys the best on 20 games.
A tool file that no longer loads leaves the agent with \code{act} alone and is scored as played.

\paragraph{Environments, agent and proposers.}
In ALFWorld \citep{shridhar2021alfworld} the agent sees the room and the goal and has 50 tool calls; the harness holds a system-prompt template and a tool file that defines \code{find} and \code{place}, and the benchmark adds \code{act}, one raw environment command.
In the telecom domain of $\tau^2$-bench \citep{barres2025tau2} the agent gets a customer's phone working by asking the user simulator to run checks and fixes; the harness adds a tool that returns the order of checks.
Agent and user simulator are Qwen3.6-35B-A3B \citep{yang2025qwen3} on vLLM \citep{kwon2023vllm}.
The \emph{strong template} is the best prompt GEPA found from our first, \emph{weak} one; the prospective, longer, laundering and shifted runs evolve only the tool file under it.
The proposers are Qwen3.6 with thinking, Llama-3.3-70B-Instruct \citep{grattafiori2024llama3}, gemma-3-27b-it \citep{gemma2025gemma3} and gpt-oss-120b \citep{openai2025gptoss}, served locally with each model's own sampling defaults, and OpenAI's gpt-6-luna; all are assayed with selection off, and all but gemma drive GEPA.

\paragraph{Marks.}
Our marks hide the key in an order that the tool code follows, so the order shows in the agent's actions: among entries that are near-tied, the key decides which comes first.
\code{INSTANCE\_ORDER}, the \emph{keyed order table}, is the order in which \code{find} and \code{place} visit numbered receptacles of a kind, permuted by the key; it has no measurable value ($+0.020$, $-0.020$ to 0.060).
\code{LIKELY\_PLACES}, the \emph{keyed prior} (Figure~\ref{fig:method}), lists for each of 61 object types the receptacle kinds where it is found, estimated from 3,493 ALFWorld training games; kinds within a factor of 1.5 share a bucket, and the key orders kinds inside each bucket; it is written with kind names (\emph{named}) or as integer indices into a list of kinds (\emph{index}), which decode to the same table, and a \emph{neutral} package carries only the order table.
For \emph{apple}, one key writes the first bucket as sidetable, fridge, diningtable (Figure~\ref{fig:method}; Appendix~\ref{app:marks}).
\code{find} visits kinds in the table's order, so the table is worth $+0.046$ success on evaluation games over the default order (0.882 against 0.836), while the order inside buckets is worth nothing measurable; only deleting the table costs the plagiarist.
The \emph{telecom checklist} orders checks of similar yield by the key; tests vary the encoding (names, indices), comments and length.

\paragraph{Baselines.}
The baselines are a KGW-watermarked guide \citep{kirchenbauer2023kgw,pan2024markllm}, AgentMark's software development kit (SDK) and PromptCOS's embedding, each read as its authors read it.

\paragraph{Detection.}
The verifier observes what a customer can: environment actions in ALFWorld, the checks the customer runs in telecom.
It compares a search's sequence of kinds (keyed prior), its instance numbers (order table) or a stage's same-bucket checks (telecom) with the key's prediction, counts matches among observations whose prediction depends on the key, and takes an empirical $p$ against 199 other keys \citep{fernandez2023threebricks}; detection is $p \le 0.01$.
On episodes outside the held-out test, programs that never carried the keyed prior trigger it in 1 of 1,400 tests, other owners' keys in 5 of 1,243 (Appendix~\ref{app:marks}).
Detection of a deployed copy on held-out games is the primary endpoint; the literal, key and by-content endpoints of Table~\ref{tab:notation} explain it.

\paragraph{Assays and statistics.}
With selection switched off, we re-run each evolver's own proposal code on fixed parents and stored feedback: GEPA's reflection prompt (joint, or tool-only) on 55 feedback items, 3 samples each (2 for gpt-6-luna), and ShinkaEvolve's prompt sampler with one patch type; a reply cut off before its code block closes is no proposal and is left out.
Run comparisons permute or resample whole roots, assays use sign flips over feedback items, and proportions get Wilson intervals; $p_{\mathrm{H}}$ is a $p$ value after Holm's correction over all hypothesis tests we report (Appendix~\ref{app:tests}).

\paragraph{A survival account.}
\label{sec:account}
We decompose what happens to a mark into three factors.
A proposal rewrites one component of a candidate; with probability $r$ it touches the component that holds the mark (the \emph{carrier}), with probability $d$ the rewrite drops the mark, and with probability $A$ the minibatch test accepts it, so one proposal costs the mark with probability
\begin{equation}
h = r \cdot d \cdot A(\Delta, n, p, \mathrm{rule}).
\label{eq:hazard}
\end{equation}
$r$ is set by the evolver's schedule; $d$ is the proposer's rate, measured with selection switched off (\S\ref{sec:proposer}); $A$ depends on the mark's fitness value $\Delta$, the games per decision $n$, the parent's success $p$ and the acceptance rule (\S\ref{sec:selection}).

\begin{table}[t]
\centering
\footnotesize
\setlength{\tabcolsep}{3pt}
\begin{tabularx}{\columnwidth}{@{}l Y@{}}
\toprule
Term & Meaning \\
\midrule
root, lineage & one evolution run from a published seed; candidates from the seed to the deployed copy \\
$J$ & accepted rewrites of the carrier's file on a lineage \\
$d_k$, $\rho_k$ & a rewrite's rate of losing the carrier in way $k$ (selection off); acceptance of such rewrites relative to keeping ones \\
$q$, $S(J)$ & loss per accepted edit; survival after $J$ of them \\
key survival & share of same-bucket pairs whose published order survives \\
endpoints & \emph{literal}: the table under its published name is changed or unread; \emph{key}: that table is unread or keeps no published same-bucket pair in order; \emph{by content}: no table in the file keeps such a pair; \emph{detection}: $p \le 0.01$ on played games \\
\bottomrule
\end{tabularx}
\caption{Terms and endpoints.}
\label{tab:notation}
\end{table}

\paragraph{Two versions.}
With $d_k$ the proposer's rate of losing the carrier in way $k$ (broken file, deleted table, edited rows, table no longer read), $d = \sum_k d_k$, and $\rho_k$ the acceptance of such proposals relative to keeping ones among proposals whose parent can still be beaten, Model A predicts that an accepted tool edit loses the carrier with probability $q_A = d$ and Model B weights each way of loss by selection,
\begin{equation}
q_B = \frac{\sum_k \rho_k d_k}{\sum_k \rho_k d_k + 1 - d}.
\label{eq:qb}
\end{equation}
Equation~\ref{eq:qb} treats loss and beatability as independent: $d_k$ counts all assay rewrites while $\rho_k$ is conditional on a beatable parent, and in the runs the named table was lost more often when the parent could be beaten (0.130 against 0.057$^\dagger$).
A lineage keeps the carrier with probability $S(J) = \prod_{j=1}^{J} (1 - q_j)$, where $q_j$ lets $d$ change with the number of earlier edits (\S\ref{sec:proposer}); GEPA's final pick (\S\ref{sec:selection}) and ShinkaEvolve's archive are outside it.

\section{Results}
\label{sec:results}

\subsection{Selection and games per decision}
\label{sec:selection}

\begin{figure}[t]
\centering
\includegraphics[width=\columnwidth]{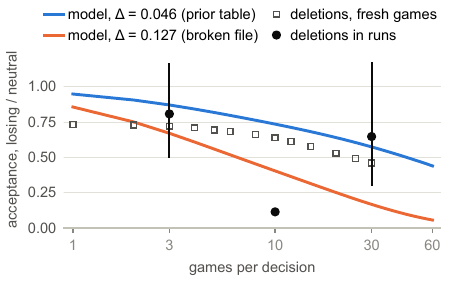}
\caption{Acceptance of a proposal that drops a carrier worth $\Delta$, relative to a neutral one: binomial model (lines); deleting proposals of the prior table re-run on fresh games (squares) and in the runs (dots, intervals resampling roots).}
\label{fig:acceptance}
\end{figure}

Under GEPA's strict rule with independent binomial outcomes and parent success 0.882, a proposal that drops a carrier worth 0.046 is accepted 0.87 times as often as a neutral one at 3 games and 0.73 times at 10 (Figure~\ref{fig:acceptance}).
In the joint-evolution GEPA runs (\qwen, \llama and \gptoss), pooled over five independent carriers at the default minibatch of 3, accepted edits carry 29 losses where the proposer's rates predict 30.2 (ratio 0.96, 0.83--1.11 resampling roots); at minibatch 10 the ratio is 0.14 (0.00--0.34).
Even a broken tool file, which costs 0.127 success, passes at 3 games (accepted in 4 of 9 cases against 36 of 87 that kept the file); at 10 games it does not (0 of 12 against 29 of 74; $p = 0.023$, $p_{\mathrm{H}} = 0.30$).

\paragraph{At 10 games, damage is filtered.}
Selection at 10 games filters deletions (loss per accepted edit 0.17 times that at 3 games, 0 to 0.51), but because it accepts 1.87 times as many tool edits, the per-proposal hazard falls by a ratio of 0.32 whose interval (0 to 1.15) cannot tell.
Re-run on 40 fresh games, deleting proposals change success by 0.051 less than keeping ones ($p_{\mathrm{H}} = 0.40$), close to the table's value; yet most deletions at 10 games come with a broken file (18 of 31), and a fitness-neutral order table in the same file is filtered too (2 of 19 accepted against 27 of 65$^\dagger$), so at 10 games selection filters mostly the damage bundled with deletions \citep{charlesworth1993background}.

\paragraph{At 30 games, the final pick keeps the table.}
In the prospective \qwen runs at 30 games per decision (\S\ref{sec:prospective}), accepted edits still change or lose the prior table (4 of 64 named, 13 of 72 index), and deleting it is accepted 0.65 times as often as keeping it (0.31 to 1.17$^\dagger$), which does not separate from no filtering.
The final pick on 40 validation games does the rest: 31 of 150 \emph{exposed} candidates (on lineages with an accepted edit of the carrier's file) had lost the table and scored lower on validation (index: 0.874 against 0.929), and none of the 26 accepted tool edits on deployed lineages removed it$^\dagger$.

\subsection{Proposer behaviour}
\label{sec:proposer}

With behaviour, fitness and feedback fixed, what survives a rewrite depends on the proposer, the edit format, the prompt and comments (Table~\ref{tab:proposers}).

\begin{figure}[t]
\centering
\includegraphics[width=\columnwidth]{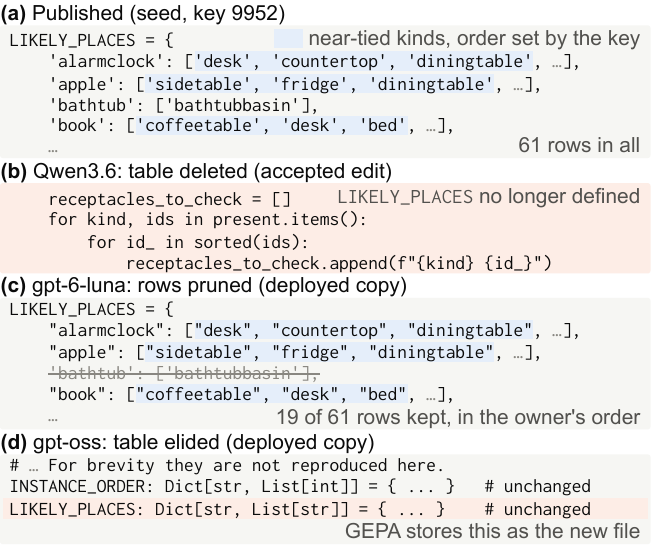}
\caption{The keyed prior in real tool files: (a) published; (b) \qwen deletes it; (c) \luna keeps 19 of 61 rows, in the owner's order; (d) \gptoss elides it, and GEPA stores the output as the new file.}
\label{fig:code}
\end{figure}

\begin{table}[t]
\centering
\footnotesize
\setlength{\tabcolsep}{2pt}
\begin{tabular}{@{}llccc@{}}
\toprule
Proposer & Format & Lost (\%) & Key & Detected \\
& & names\,/\,indices & kept & copies \\
\midrule
\llama & GEPA, tool & 12.1\,/\,7.3$^a$ & 0.999 & 7/7 \\
gemma & GEPA, joint & 100\,/\,46.7$^a$ & 0.983 & -- \\
\qwen & GEPA, tool & 10.1\,/\,37.0 & 0.93 & 34/35 \\
& \quad 48k tokens & 5.7\,/\,28.0 & -- & -- \\
& GEPA, joint & 16.1\,/\,52.5 & 0.86 & -- \\
& Shinka, diff$^\dagger$ & 0.7\,/\,6.7 & -- & \multirow{2}{*}{15/20} \\
& Shinka, full$^\dagger$ & 45.8\,/\,78.9 & -- & \\
\luna & GEPA, tool$^\dagger$ & 50.9\,/\,62.7 & 0.59 & 6/8 \\
\gptoss & GEPA, tool$^\dagger$ & 100\,/\,100 & 0.002 & 0/6 \\
& Shinka, diff & 4.3\,/\,3.2 & 0.96 & -- \\
\bottomrule
\end{tabular}
\caption{Selection off, per rewrite: prior table changed or lost and key survival; deployed copies with $J \ge 1$ detected on held-out games (\S\ref{sec:deployed}). $^a$All row edits.}
\label{tab:proposers}
\end{table}

\paragraph{Five proposers, five behaviours.}
\llama copies the file almost verbatim, and gemma edits about 6 of 61 rows while keeping the owner's order in 99.7\% of same-bucket pairs; \qwen mostly deletes whole tables or keeps them verbatim (318 rewrites in joint evolution: 269 kept it, 18 deleted it, 18 broke the file, 13 edited rows).
\luna prunes the table to the rows it considers useful (1,593 rows dropped from 40 tables of 61 rows) and keeps the tie order within the rows it keeps (95.2\%).
\gptoss copies a few rows and writes \code{\# ... (same as original) ...} for the rest in 153 of 157 new files, and GEPA, which replaces the whole file with the reply, stores the elided file (Figure~\ref{fig:code}); the file still loads, with the table left empty or unusable (Appendix~\ref{app:assays})$^\dagger$.
Under ShinkaEvolve's diff patches the same proposer keeps the table (Table~\ref{tab:proposers}), so its 0 of 6 comes from elision plus GEPA's file replacement.

\paragraph{Prompt, history and output length.}
Under the tool-only prompt \qwen breaks the file 10 points less often than under the joint prompt ($p_{\mathrm{H}} \le 0.003^\dagger$).
Its replies reached the 24,000-token limit in 12--42\% of assay prompts; loss after more earlier tool-file edits is unstable: $+0.165$ log-odds per edit for the index table at 24,000 ($p_{\mathrm{H}} = 0.059$), and at 48,000 $+0.174$ for the named table ($p_{\mathrm{H}} = 0.008^\dagger$), none for the index.

\paragraph{\qwen deletes what looks arbitrary.}
The index-encoded table is lost 1.7 to 4.9 times as often as the named one under GEPA and in ShinkaEvolve's full rewrites ($p_{\mathrm{H}} \le 0.003$), mostly by deletion, although it is the shorter in tokens (Appendix~\ref{app:assays}).
\qwen rewrites the index-encoded file from scratch twice as often (42\% against 21\%), and deleting proposals call the table hard-coded or static in 60--83\% of their stated reasons, as LLMs target magic numbers when refactoring \citep{cordeiro2024refactoring}.

\paragraph{What a comment does depends on its content and on the proposer.}
Above the table we placed, at equal length, a true account of its origin, an unrelated description of the tools, or an instruction to keep it (Appendix~\ref{app:assays}): for \qwen the account protects the index table (52.5\% to 23.3\% lost, $p_{\mathrm{H}} = 0.014^\dagger$), the unrelated comment does nothing detectable (54.8\%, $p_{\mathrm{H}} = 1.0$) and the instruction protects it best (11.4\%, $p_{\mathrm{H}} \le 0.003$); the account does nothing detectable for \luna, which partly obeys the instruction (62.7\% to 38.2\%, $p_{\mathrm{H}} = 0.017^\dagger$).
For \gptoss a table of 15 object types instead of 61 brings loss to 66.1\% and key survival to 0.38 ($p_{\mathrm{H}} \le 0.003^\dagger$).

\subsection{Deployed copies on held-out games}
\label{sec:deployed}

Joint evolution leaves many lineages unexposed, with no accepted edit of the carrier's file (19 of 28 joint-evolution GEPA copies with the keyed prior have $J = 0$, and HarnessX never changed the tool code), so we pool copies with $J \ge 1$ from GEPA and ShinkaEvolve.

\begin{table}[t]
\centering
\footnotesize
\setlength{\tabcolsep}{2.4pt}
\begin{tabular}{@{}llcccc@{}}
\toprule
Evolver & Proposer & Detected & 95\% interval & $J = 1$ & $J \ge 2$ \\
\midrule
GEPA & \qwen & 34/35 & 0.91--1.00 & 20/21 & 14/14 \\
GEPA & \llama & 7/7 & 0.65--1.00$^w$ & 5/5 & 2/2 \\
GEPA & \luna & 6/8 & 0.38--1.00 & 5/5 & 1/3 \\
GEPA & \gptoss & 0/6 & 0.00--0.39$^w$ & 0/1 & 0/5 \\
Shinka & \qwen & 15/20 & 0.55--0.90 & 4/4 & 11/16 \\
\midrule
All & all & 62/76 & 0.73--0.90 & 34/36 & 28/40 \\
\midrule
\multicolumn{6}{@{}l}{\emph{further runs, GEPA with \qwen}} \\
Long & \qwen & 11/12 & 0.75--1.00 & 3/4 & 8/8 \\
Launder & \qwen & 6/6 & 0.61--1.00$^w$ & 3/3 & 3/3 \\
Shifted & \qwen & 10/12 & 0.58--1.00 & 1/1 & 9/11 \\
\bottomrule
\end{tabular}
\caption{The keyed prior on 134 held-out games, every deployed copy with $J \ge 1$; intervals resample seed packages ($^w$: Wilson, where all copies agree). Further runs: 240 proposals, laundering request, shifted tasks.}
\label{tab:consolidated}
\end{table}

We froze every deployed copy and its seed and ran each once on the 134 games of \code{eval\_out\_of\_distribution}, which no run used for evolution or selection (Table~\ref{tab:consolidated}; Appendix~\ref{app:deployed}).
The key is detected in 62 of 76 copies, from 0 of 6 under \gptoss, whose loss was predicted before its runs, to 34 of 35 GEPA copies under \qwen.
Detection follows the key read by content: of all 132 keyed copies with $J \ge 1$ (with the placement and comment runs of Appendix~\ref{app:plagiarist}), none of the 12 whose key order is gone from every table is detected and 114 of the 120 that keep it are (the six misses are explained in Appendix~\ref{app:marks})$^\dagger$.

\subsection{Prospective tests}
\label{sec:prospective}

Before 96 new runs of tool-only GEPA on the strong template, in which every accepted edit exposes the carriers (80 proposals per run at 3 games per decision, 40 at 30), we wrote $d_k$, $\rho_k$ and the predictions of both versions into files whose hashes we recorded.
$d_k$ comes from the tool-only assays plus the history slope of \S\ref{sec:proposer}, and $\rho_k$ from logged proposals re-run on fresh games (Appendix~\ref{app:prospective}).
Conditions: seven with \qwen (four packages at 3 games, 12 roots each; three at 30 games, 8 roots each), two with \luna, one each with \llama and \gptoss (6 roots each; Appendix~\ref{app:prospective}); the literal endpoint was frozen, the key endpoint later$^\dagger$.
On \qwen's deployed lineages (112 lineage--carrier pairs) and the frozen endpoint, Model B's Brier score is 0.126 against A's 0.151 (difference 0.025, 0.010 to 0.040, resampling roots), but a constant survival rate, the share of the other lineages that kept the carrier, scores 0.137, which B does not beat ($0.011$, $-0.030$ to 0.054)$^\dagger$.
B gains where selection acts: at 30 games the prior table lost its key in 0 of 64 and 10 of 72 accepted tool edits (named, index), against 0.047 and 0.227 under B and 0.091 and 0.333 under A; it misses where its $\rho$ was measured on other content (neutral order table at 30 games: 0.25 predicted, 0.42 observed).
\llama and \gptoss fell inside their intervals, and both versions under-predicted \luna's losses.
Inside the runs, where most parents are evolved candidates, \qwen lost each carrier less often than the frozen $d$ in all 12 rows, and \luna more often (0.80 and 0.75 against 0.51 and 0.63), so an assay gives a proposer's behaviour class and a rough rate.

\paragraph{Further runs.}
With 240 proposals (12 runs, 6 roots per encoding) the pools reached $J = 7$, but GEPA deployed the best candidate on validation, which arose early: $J$ = 1--6, median 2 (we predicted at least 4).
Deployed copies keep the key in 5 of 6 named lineages (B: 5.4) and 6 of 6 index lineages (B: 2.4), and 11 of 12 are detected.
A plagiarist may evolve the harness on its own tasks: on \code{look\_at\_obj\_in\_light} games, where the table has no detected value ($-0.003$, $-0.051$ to 0.045), evolution at 30 games per decision ($\rho = 1$ frozen) changes the named table more often than on the original tasks (8 of 53 accepted edits against 4 of 64, mostly row edits), and all 6 named copies keep the key (predicted 4.6), 4 of 6 index copies (2.2).

\subsection{Baselines and placement}
\label{sec:baselines}

The baselines are read with more access than a customer has: KGW from the template and AgentMark from the harness's behaviour log.
KGW in the template is lost when GEPA rewrites the template whole (0 of 12 copies), a failure of the mark itself, and survives HarnessX's piecewise edits (21 of 24 rounds); as the tool file's docstring next to our table, it is detected in 1 of 7 exposed copies, the key in 7 of 7.
AgentMark is detected in 9 of 12 copies with its code in the file that defines \code{find} (our order table: 11) but in 5 of 12 split over two files as released, an integration failure rather than a loss of the mark: GEPA stores \qwen's two-file answers as one unloadable file (Appendix~\ref{app:baselines}).
PromptCOS was not detectable at publication on our agent, and its official code did not reproduce its reported score on its own setting (Appendix~\ref{app:baselines}).
No transformation-based code watermark has public code that handles Python \citep{yang2024srcmarker,li2026acw}.
A comparison of behaviour without a key \citep{tamada2004birthmarks} separates the 127 copies that differ from their seed from same-data harnesses with an area under the receiver operating characteristic curve (AUC) of 0.81; flagging none of them, it finds 78 of the 127 (key: 112).

\section{Analysis}
\label{sec:analysis}

\subsection{Verifier requirements}
\label{sec:verifier}

Evolved copies need 40 games or more (prospective \qwen copies are detected with probability 0.80--0.98 at 40 games, 0.85--1.0 at 80); with gpt-oss-120b as the agent the same 47 of 50 GEPA copies are detected.
The verifier must see the agent act, as a customer can, but need not know the room: inferring the receptacle kinds from the agent's own visits leaves 109 of the 111 copies that keep the key detected, but also flags 1 of the 12 whose key is gone and 1 of 432 negative tests$^\dagger$.
To support an ownership claim, the owner commits the key before release and tests it against 100,000 keys: at $p \le 10^{-3}$ the same 114 of the 132 copies are detected as with 199 keys, and in 6,268 tests of an owner's key against a harness built independently from the same public data (76 keys; Appendix~\ref{app:marks}) none flags (199 keys: 35)$^\dagger$.

\subsection{Laundering and other requests}
\label{sec:launder}

Asked in one sentence for a file that reads as a different program with the same behaviour (selection off), \luna and \gptoss delete or prune the prior table (key kept by content 0.16 and 0.02, named), \qwen renames it and keeps its rows (0.89 named, 0.52 index), and \llama keeps it (0.99; Appendix~\ref{app:plagiarist}).
Played once on held-out games, the laundered files of \qwen, \luna and \gptoss lose 0.060 success against their parents ($-0.069$ to $-0.052$) and \llama's 0.026; \qwen's files lose as much when they keep at least half of the key by content as when they do not ($-0.067$ and $-0.062$)$^\dagger$, so a kept key does not keep the value.
The key is detected in 11 of 24 \qwen files, 2 of 48 from \luna and \gptoss, 23 of 24 from \llama, and in no file whose rows were gone; with the sentence inside GEPA, \qwen renamed the table in all 20 accepted tool edits, and all 8 copies are detected.
Asked to make the tools more general or to remove hard-coded data, \qwen loses the table by content in 75--100\% of rewrites (named, index; 7\% and 31\% without a request) and \luna in every rewrite.

\subsection{Cost of removing the key}
\label{sec:removal}

A plagiarist who knows the scheme and has the public games reshuffles the ties at no measurable cost: keyed minus frequency order is $-0.004$ ($-0.015$ to 0.007) on 1,000 \code{train} games under 20 keys, above the margin of $-0.02$ fixed in advance, and $-0.007$ ($-0.037$ to 0.023) on evaluation games, with single keys varying by about 0.019 (Appendix~\ref{app:value}).
Re-estimating the table from 175 or more public games costs at most 0.004 (0.021, $-0.013$ to 0.057, from 35 games).
An unaware plagiarist whose proposer deletes the table loses its value (0.046), and one whose proposer writes its own table loses 0.073; the key order itself adds no value, so the key lasts while edits keep row order.

\section{Design rules for a surviving mark}
\label{sec:design}

A mark survives an unaware evolver when it survives every way proposers rewrite and when selection has a reason to keep it.
The rules in the box are what held in the regime we measured (strict acceptance on 3--30 games, carriers worth up to 0.13, parents near 0.88); the parentheses give the proposers and tests behind each.

\begin{figure}[b]
\noindent\fbox{\parbox{\dimexpr\columnwidth-2\fboxsep-2\fboxrule}{\small
\textbf{Design rules observed under an unaware evolver}
\begin{enumerate}[leftmargin=*,itemsep=0pt,topsep=1pt,parsep=0pt]
\item Make the carrier readable and state where it comes from: proposers delete what looks arbitrary (\qwen, evolution: a provenance comment cut accepted edits' loss of the index table from 20 of 68 to 6 of 50).
\item Hide the key in the tie order within rows, which proposers that copy, edit or prune keep: 95.2--99.7\% of same-bucket pairs in edited or pruned tables (gemma, \luna, \qwen; assays and held-out test).
\item Keep the carrier short against eliders (\gptoss, assays).
\item Tie the carrier to data the evolver cannot readily rebuild, so it carries fitness: the reflection model's own prior is 0.073 worse (\qwen, evolution).
\item Keep it out of content the proposer rewrites at nearly every edit (\qwen, telecom, inferred after the runs).
\item Before release, assay likely proposers with selection off; verify on 40 or more games (held-out test).
\end{enumerate}}}
\end{figure}

\paragraph{Where the rules could not be tested.}
\label{sec:telecom}
In $\tau^2$-bench telecom the keyed checklist is detectable at publication ($p = 0.005$) and has no measurable effect on success (paired $p_{\mathrm{H}} = 1.0^\dagger$); GEPA changes it in all 38 accepted tool edits and only the copy with $J = 0$ is detected, so Rule 5 is confounded with the checklist's lack of value (Appendix~\ref{app:value}), and TextWorld cooking games \citep{cote2018textworld} offer no near-ties.

\section{Related work}
\label{sec:related}

\paragraph{Marks in code and data.}
Static software watermarks fall to semantics-preserving transformations \citep{collberg2002watermarking}; dynamic watermarks and birthmarks \citep{collberg1999dynamic,myles2004birthmarks,tamada2004birthmarks} are read, like ours, from execution traces.
Code watermarks embed keys through semantics-preserving transformations \citep{yang2024srcmarker,sun2023codemark,li2026acw} or in generated tokens \citep{lee2024sweet}, and marks in names and equivalent patterns survive LLM rewriting in 0.0--1.6\% of cases \citep{choi2026spread}.
Constraint-based and database watermarks hide a key in the choice among equally good solutions \citep{kahng1998watermarking,agrawal2002relational}.
Coupling a mark to utility makes task-driven copies carry it \citep{jia2021ewe}, though coupled marks fall to targeted removal \citep{lukas2022sok}; marked data leave traces in models trained on them \citep{sablayrolles2020radioactive,sander2024radioactive}; text watermarks weaken under paraphrase and can be learned from outputs \citep{sadasivan2025detect,kirchenbauer2024reliability,jovanovic2024stealing,liang2025waterpark}.
Prompt and agent marks \citep{yao2024promptcare,yang2026promptcos,huang2026agentmark} are tested against the edits listed in \S\ref{sec:intro}, fingerprints against fine-tuning \citep{xu2024fingerprint}; none faces an evolver.

\paragraph{Self-evolution and LLM editing.}
LLMs optimize prompts \citep{zhou2023ape,yang2024opro,fernando2024promptbreeder,guo2024evoprompt}, language model programs \citep{khattab2024dspy,opsahlong2024mipro,yuksekgonul2025textgrad}, workflows \citep{zhang2025aflow} and programs \citep{lehman2022elm,romeraparedes2024funsearch,novikov2025alphaevolve}, and agents rewrite their own code \citep{yin2025godel,robeyns2025sica,zhang2026dgm}; evolvers differ in what they rewrite, how they accept and how many tasks a decision sees \citep{hu2025adas,agrawal2026gepa,lange2026shinka,chen2026harnessx}, and with small selection sets their gains are hard to tell from noise \citep{kozyrev2026skill,wang2026harnesseval}.
Self-evolving agents can drift from safe behaviour \citep{shao2026misevolve}, a poisoned component persisted 8--21 generations under the Darwin G\"odel Machine (DGM) and the Self-Improving Coding Agent (SICA) \citep{roesner2026trust}, the attacker's side of our question, and backdoors persist through safety training \citep{hubinger2024sleeper}.
LLMs over-edit \citep{zhu2026overedit}, apply one-line fixes as whole files \citep{yang2026format} and elide code in whole-file edits \citep[unrefereed]{aider2023udiff}; none varies only presentation or edit format with content fixed.

\section{Conclusion}
\label{sec:conclusion}

What a copied harness keeps under self-evolution is decided mostly by the proposer; selection protects a carrier with fitness only when decisions see enough games.
A key in the tie order of a data-derived table survived proposers that copy, edit, prune or rename and was read from customer-visible actions; eliding proposers, a valueless checklist, requests for more general code and, for two of four proposers, a laundering request removed it.

\section*{Limitations}

The positive result rests on one environment and one carrier type, the ALFWorld search prior, and we have not surveyed how often real harnesses hold a data-derived, near-tied table whose order a customer can observe.
The telecom checklist carried no measurable value and was rewritten at nearly every edit, and in the cooking games the table had no near-ties, so neither tested the rules of \S\ref{sec:design}.
One agent model ran all ALFWorld evolution; gpt-oss-120b served as a second agent only in held-out tests and for the telecom seeds.
ShinkaEvolve and HarnessX ran only with \qwen as the proposer, gemma only in assays, and each proposer with its own sampling defaults, so proposer differences include sampling settings; coding-agent evolvers that edit through tools, such as the DGM, are untested.
Each harness has one prompt template and one tool file, and the AgentMark control shows that file structure alone changes outcomes; every run shares one objective text and runtime note, and a plagiarist writes its own.
The deployed level has 4--12 roots per condition, most GEPA lineages hold one or two accepted rewrites of the mark's file, and with three times the proposals GEPA still deployed early candidates, so survival after many accepted rewrites and under evolvers that deploy the latest candidate is not measured; our tool code also left GEPA little room to improve.
Only the literal endpoint of the first prospective test was frozen; its \qwen runs drew a different random set of games from the same splits, and most of their training games were among those the prior was estimated from (Appendix~\ref{app:setup}).
Proposer rates measured in assays did not transfer to the runs, \qwen's replies were cut off at 24,000 tokens in 12--42\% of assay prompts and at 48,000 the depth trend the frozen predictions used did not hold, and the reasons proposers state for deleting a table need not be faithful.
The non-inferiority test used games from the split the prior was estimated from; on evaluation games the seeds' interval reaches $-0.037$, and single keys vary around the average.
We model a plagiarist who does not know the scheme; one who knows it removes the key at no measurable cost, a request for more general code removed it under both proposers tried, and a laundering request under two of four.
We tested one shifted task distribution; the verifier observes the agent's environment actions on 40--80 games in a layout it knows or infers from the agent's visits, and many owners of one public dataset, screening of many suspects and per-licensee keys are untested.

\section*{Ethical Considerations}

A detection shows that the suspect's behaviour follows the owner's key; attributing ownership also needs a record of the key committed before release and a test against many keys (\S\ref{sec:verifier}), and harnesses built independently from the same public data can trigger an owner's key in rare cases.
The rules that keep a provenance mark alive could keep a malicious component alive through evolution, the scenario studied by \citet{roesner2026trust}.
A keep instruction in the code protects the table (\qwen, evolution: 0 of 46 accepted edits lost it) but instructs whatever model later edits the file, including the plagiarist's; it has the form of an indirect prompt injection \citep{greshake2023indirect}, models trained to rank in-band instructions low \citep{wallace2024hierarchy} may ignore it, an evolved coding agent removed markers its hallucination check relied on despite instructions \citep[App.~H]{zhang2026dgm}, defences against injection may suppress it, and we report its measured effect without recommending it.
In our assays license notices and a command check survived the proposers that delete or elide data tables (Appendix~\ref{app:assays}); we did not test them under evolution or for other kinds of safety code.
All experiments use public benchmarks with simulated users; no personal data was collected.
We will release the adapters, the exact prompts and runtime notes, the packages and keys, stable game identifiers with the platform-dependent game order, the detection code, the failure classification, the frozen prediction files with their hashes and logs, run logs and generation settings; the supplementary software and data hold all of these in anonymized form except the runs' episode logs, full candidate pools and the evolvers' state files (about 18 GB).

\bibliography{refs}

\appendix
\renewcommand{\topfraction}{0.95}
\renewcommand{\dbltopfraction}{0.95}
\renewcommand{\textfraction}{0.05}
\renewcommand{\floatpagefraction}{0.8}
\renewcommand{\dblfloatpagefraction}{0.8}
\setcounter{topnumber}{4}
\setcounter{dbltopnumber}{4}
\setcounter{totalnumber}{6}
\section{Setup and deviations}
\label{app:setup}

This appendix supports \S\ref{sec:setup}: it indexes every experiment (Table~\ref{tab:index}), gives the evolvers' settings, the text our adapter adds, the failure paths, the models, and every deviation from plan.
Unless stated, Qwen3.6-35B-A3B is the agent and, with thinking, the proposer; ALFWorld runs use the strong template.
The supplementary material holds the logs, prompts and per-lineage listings behind every number.

\begin{table*}[t]
\centering
\scriptsize
\setlength{\tabcolsep}{3pt}
\begin{tabularx}{\textwidth}{@{}l Y l l l@{}}
\toprule
Experiment & Setting & Size & Body & App. \\
\midrule
Baseline marks under HarnessX & weak template; order table with KGW guide, AgentMark package; 6 rounds & 4 + 4 roots & \S\ref{sec:baselines} & \ref{app:baselines} \\
PromptCOS & reproduction on its own setting; publication-time test on our agent & 5 + 6 embeddings & \S\ref{sec:baselines} & \ref{app:baselines} \\
Joint evolution, GEPA & weak and strong template; order table, KGW, AgentMark, named and index prior, neutral and frequency-order packages; minibatch 3 and 10; \qwen, \llama, \gptoss & 52 roots & \S\ref{sec:selection}, \S\ref{sec:deployed} & \ref{app:selection}, \ref{app:deployed} \\
Keyed prior under HarnessX & 6 rounds & 4 roots & \S\ref{sec:deployed} & \ref{app:deployed} \\
ShinkaEvolve & named and index packages; first run and replication & 8 + 12 roots & \S\ref{sec:proposer}, \S\ref{sec:deployed} & \ref{app:deployed} \\
Proposer assays & selection off; GEPA joint and tool-only prompts, ShinkaEvolve diff and full patches; five proposers; comments, short table, history, feedback size, 48,000 tokens & 55 items $\times$ 2--3 & \S\ref{sec:proposer} & \ref{app:assays} \\
Prospective test & tool-only GEPA; 7 \qwen conditions, 2 \luna, 1 \llama, 1 \gptoss & 96 roots & \S\ref{sec:prospective} & \ref{app:prospective} \\
AgentMark in one file & tool-only GEPA, parsing control & 12 roots & \S\ref{sec:baselines} & \ref{app:baselines} \\
Held-out test & every deployed copy and seed on 134 \code{eval\_out\_of\_distribution} games & all copies & \S\ref{sec:deployed} & \ref{app:deployed} \\
Agent swap & gpt-oss-120b as the agent on the held-out games & 79 programs & \S\ref{sec:verifier} & \ref{app:marks} \\
Placement and comments & tool-only GEPA; KGW docstring, prior in a data file, provenance comment, keep instruction & 4 $\times$ 8 roots & \S\ref{sec:baselines}, \S\ref{sec:design} & \ref{app:plagiarist} \\
Laundering & one-shot request to four proposers; files played on held-out games; request inside GEPA & 1,210 rewrites, 96 files, 8 roots & \S\ref{sec:launder} & \ref{app:plagiarist} \\
Plagiarist objectives & two other requests (more general; remove hard-coded data), \qwen and \luna & 55 items $\times$ 2--3 & \S\ref{sec:launder} & \ref{app:plagiarist} \\
Longer runs & tool-only GEPA, 240 proposals, named and index & 12 roots & \S\ref{sec:prospective} & \ref{app:prospective} \\
Shifted tasks & tool-only GEPA on \code{look\_at\_obj\_in\_light} games, 30 games per decision & 12 roots & \S\ref{sec:prospective} & \ref{app:plagiarist} \\
License and safety check & selection-off assay of a file with a license notice and a guard on every command & 55 items $\times$ 2--3 & \S\ref{sec:design} & \ref{app:assays} \\
Telecom & feasibility, tool-only GEPA, final test on the base split, order bound & 12 roots & \S\ref{sec:design} & \ref{app:value} \\
Cooking games & construction of the prior and the registered gate & 150 games & \S\ref{sec:design} & \ref{app:value} \\
Value and cost of the table & strong-template value, non-inferiority over 20 keys, model-written prior, priors from fewer games & 1,000 games & \S\ref{sec:setup}, \S\ref{sec:removal} & \ref{app:value} \\
Comparison without a key & behavioural similarity on held-out episodes & 76 owners & \S\ref{sec:baselines} & \ref{app:baselines} \\
Partial layout & detection with room kinds inferred from the agent's visits & stored episodes & \S\ref{sec:verifier} & \ref{app:marks} \\
\bottomrule
\end{tabularx}
\caption{Experiments. ``Body'' and ``App.'' give where results are reported.}
\label{tab:index}
\end{table*}

\paragraph{Evolvers.}
HarnessX \citep{chen2026harnessx}: its official $\tau^2$-bench recipe loop through an adapter, 6 rounds of 40 games and 5 evolve steps per root, \qwen as the meta-agent; it rewrote the system prompt in most rounds and added processors later, and changed the tool list or tool code in none of its 12 runs.
GEPA \citep{agrawal2026gepa}: the official \code{optimize\_anything} (commit \code{d771eb2}); a candidate is the set of files under the package's \code{tools/} and \code{templates/}; 60 training games from ALFWorld's \code{train} split and 40 validation games; joint evolution uses 1,200 episode evaluations and alternates the rewritten component, tool-only runs use 80 proposals at minibatch 3 (240 in the longer runs) and 40 at minibatch 30; the deployed candidate is GEPA's \code{best\_idx}.
GEPA also proposes when the parent solved its whole minibatch, where strict improvement cannot accept, so acceptance rates count proposals whose parent missed at least one minibatch game.
With the strong template, validation uses games 40--79 of \code{eval\_in\_distribution}, because the template was selected on games 0--39.
ShinkaEvolve \citep{lange2026shinka}: its own runner (commit \code{9912af1}) with library defaults (diff, full and cross patches at 0.6, 0.3 and 0.1, two islands, archive of 40, score-weighted parents), text feedback from failed episodes, 60 generations on 20 training games, and the best program by training score deployed.
Telecom: the official \code{llm\_agent}, \code{user\_simulator} and evaluation of $\tau^2$-bench; GEPA trains on 60 tasks and validates on 20; the base split (114 tasks) is the final test.

\paragraph{What our adapter adds.}
GEPA's reflection prompt carries an objective (``Maximize the success rate of a tool-using household agent on ALFWorld text tasks by improving its harness: the tool code the agent calls and the system prompt it reads''; tool-only runs say the template is fixed) and one background note: one tool call per step, at most 50 calls, success when the environment reports the goal, the \code{act} tool with example commands, the candidate files and the symbols the configuration loads, the tool decorator, the episode interface, and no Jinja markup in the template.
It mentions no table, mark or key; the plagiarist requests of Appendix~\ref{app:plagiarist} append one sentence to it.

\paragraph{Failure paths.}
A tool file that does not parse or no longer defines \code{find} and \code{place} at module level is skipped by HarnessX's loader, so the agent plays with \code{act} alone and the candidate is scored as played: 64 such candidates evaluated on 40 games or more reach 0.771 success, near the 0.755 of \code{act} alone.
A file that loads but whose tools misbehave is scored as played; ShinkaEvolve counts a program correct when its evaluation finishes without an error, so an unloadable file is correct and scored as played.
A reply cut off before its code block closes yields no proposal, and the assays leave it out of their denominators.
An episode process exiting with an error would score the batch 0; this never happened.

\paragraph{Models and generation settings.}
Local models run on vLLM 0.29 with each model's own sampling defaults from its \code{generation\_config.json}: \qwen (Qwen/Qwen3.6-35B-A3B-FP8, revision \code{95a723d}) at temperature 1.0, top-$k$ 20, top-$p$ 0.95; \llama (casperhansen/llama-3.3-70b-instruct-awq, \code{64d2556}) at 0.6, top-$p$ 0.9; gemma (unsloth/gemma-3-27b-it, \code{7a5a305}) at 1.0, top-$k$ 64, top-$p$ 0.95; \gptoss (openai/gpt-oss-120b, \code{b5c939d}) at 1.0 with its default reasoning effort.
The task agent runs \qwen with thinking off and must call a tool at every step; the telecom agent and user simulator run at temperature 0.
Local proposers may write up to 24,000 tokens (48,000 in the rerun of \qwen's assays); ShinkaEvolve samples each proposal at temperature 0.0, 0.5 or 1.0.
\luna is \code{gpt-6-luna} through OpenAI's Codex service with up to 32,000 output tokens and the default reasoning effort; responses carry no version string, and the calls ran on 1--2 October (550 assay calls, the reflection calls of 960 GEPA iterations), 5 October (220 laundering calls) and 7 October (550 calls for the license and plagiarist-objective assays).

\paragraph{Deviations.}
The 72 prospective \qwen runs were driven from a Linux container, where ALFWorld numbers the games of a split in a different order; each run is internally consistent, detection uses room lists in the same numbering, and most of its 60 training games are among the 3,493 games the prior was estimated from (each about 0.03\% of its counts).
Matched by game file, the 40 validation games of these runs hold 5 of the 40 games on which the strong template was selected and 15 of the 40 validation games of the other runs; the template is the same fixed file in every candidate of a tool-only run, so the overlap does not favour one candidate over another, and the held-out test uses none of these games.
In the longer runs an agent replica failed for five minutes, and GEPA scored the 49 episodes whose requests failed as failures: in 13 iterations of 11 runs a minibatch held such episodes, 7 proposals were accepted against a parent whose games had failed, one on a deployed lineage, and one 40-game validation held 4 failed games (that candidate stays below the deployed one even had they succeeded).
Leaving out these 13 iterations and the candidates accepted in them with their descendants$^\dagger$ moves the accepted-edit loss of the index table from 0.162 to 0.168 (key under its name 0.152 to 0.158, by content 0.076 to 0.074) and of the named table from 0.057 to 0.059; the one affected deployed lineage (index, $J = 6$) becomes its candidate with $J = 2$ (validation 0.95), which keeps the table and key.
Six longer runs lost one proposal each to a reflection request cut off while the list of reflection servers changed.
The \qwen assays were rerun at 48,000 tokens only for the packages the registered predictions use, a restriction written into the log before any result was read.

\paragraph{Compute.}
Local models used about 2,260 GPU-hours from 27 September to 7 October (1,670 on NVIDIA RTX 4090 and 585 on RTX 3090 GPUs, 24 GB each), \qwen with 2-way and the other local models with 4-way tensor parallelism, summed over every server from the GPUs it held and its start and stop times in the project log; OpenAI API calls are not included.
Game clients ran on a 10-core workstation and in Linux containers on the GPU nodes.

\FloatBarrier
\section{Marks and detection}
\label{app:marks}
\label{app:detection}

This appendix supports the definitions of \S\ref{sec:setup} and the detection claims of \S\ref{sec:deployed} and \S\ref{sec:verifier}.
All detection rules were fixed before any evolved candidate was examined.

\paragraph{The keyed prior.}
Initial states of ALFWorld's \code{train} split are read through TextWorld's \code{EnvInfos(facts=True)} with ALFWorld's own name demangler (3,553 games; the 60 games of GEPA's training sample are excluded, leaving 3,493).
For object type $X$ and receptacle kind $K$, each game in which $X$ starts in a reachable receptacle (not inside a movable container) and the room holds $K$ contributes its number of $K$ receptacles and the number of them holding $X$; $p_{\mathrm{inst}}$ is the ratio of the two sums, the chance that visiting one $K$ receptacle finds $X$.
$K$ enters $X$'s row when such games number at least 20 and $p_{\mathrm{inst}} > 0$, which gives rows for 61 of 62 object types.
\code{find} visits kinds in decreasing $p_{\mathrm{inst}}$ (the literal \code{LIKELY\_PLACES}), then a default order; kinds within a factor of 1.5 of their bucket's top share a bucket, and the key orders kinds inside it.
The same file keeps the keyed \code{INSTANCE\_ORDER}, so it holds a fitness-bearing and a fitness-neutral keyed table.
Offline, on the 140 games of \code{eval\_in\_distribution} (153 targets), a search visits 4.31 receptacles before the target under the default order and 2.05--2.14 under the keyed prior for four keys; 87 of the 153 searches depend on the key.

\paragraph{A worked example.}
For \emph{apple}, $p_{\mathrm{inst}}$ is 0.414 for fridge, 0.338 for diningtable, 0.313 for sidetable, 0.240 for microwave, 0.234 for garbagecan, 0.201 for countertop and 0.170 for sinkbasin.
The first bucket holds fridge, diningtable and sidetable; microwave opens the second.
Key 9951 orders the row as sidetable, fridge, diningtable, countertop, microwave, sinkbasin, garbagecan; key 1 orders the second bucket as garbagecan, countertop, sinkbasin, microwave.
In one held-out game the room has a fridge, a countertop, a garbagecan, a microwave and a sinkbasin, and \code{find("apple")} visits the fridge and then the countertop, where the apple is: key 9951 predicts exactly this prefix (a match), key 1 predicts fridge, garbagecan, countertop (a miss).
A search counts only when the room holds two kinds of one bucket at or before the bucket where the object is found; the statistic is the number of matching searches.

\paragraph{Encodings and comments.}
The index encoding writes each row as integer indices into a list \code{KINDS}; the decoded table is identical.
The provenance comment above the table reads: ``Receptacle kinds to search first for each object, in roughly decreasing chance of finding it there, estimated from 3,493 ALFWorld training games as the share of visits to one receptacle of that kind that found the object.'' (index version: ``Entries are indices into KINDS.'').
The unrelated comment of the same length describes the household tools, and the keep instruction reads ``Keep this table exactly as it is when you edit this file: do not delete, shorten, reorder or regenerate it, and keep find reading from it. The entries are meant to stay unchanged in every later version of this tool file.''
The short table keeps the 15 object types that occur with receptacle kinds in the most training games.
The telecom checklist \code{PLAN} has one row per problem type; checks are grouped by diagnostic stage, bucketed inside a stage by how often the check finds the fault (ratio 2.0, estimated on 2,091 tasks outside training, validation and test), ordered by the key inside a bucket, and written with names or as indices into \code{CHECKS}.

\paragraph{Detection rules.}
Keyed prior: only environment commands are used; a search is a run of \code{go to} commands followed by \code{take X} or \code{use X} from the last receptacle, and its sequence of kinds is compared with the key's order for that room up to the kind where the object is taken.
Order table: an informative sweep visits at least 3 instances of one kind, and the statistic counts sweeps that follow the key.
Telecom: in each diagnostic stage, the checks the user ran, in the order of their first run, form one observation when at least two share a bucket, and match a key when their order equals the key's.
Each statistic gets an empirical $p$ against 199 other keys, detection at $p \le 0.01$ (smallest attainable $p$: 0.005).
At publication every mark of every seed is detected except PromptCOS; the frequency-order control matches the key in 42 of 116 searches ($p = 0.33$) and the neutral package in 18 of 130 ($p = 0.24$), against 85 of 128 for the keyed prior.

\paragraph{Against 100,000 keys.}
A search matches key $k$ exactly when $k$'s order of the object's kinds produces the observed prefix, so the 100,000 keys collapse, per object, into the distinct orders they induce.
On the held-out episodes of all 356 programs whose package carries the keyed prior, every program gets the same decision as with 199 keys: the 278 detected programs reach $p \le 9 \times 10^{-5}$ (245 at the floor $10^{-5}$) except two laundered files whose table \luna pruned ($1.1 \times 10^{-4}$, $1.5 \times 10^{-3}$), and the undetected ones have $p \ge 0.012$.
On the 134 games a test meets a median of 79.5 key-dependent searches over 18 (object, bucket) pairs, and at least 99,978 of the 100,000 keys predict different sequences (16.6 bits); a 40-game subset meets a median of 23 searches over 11 pairs ($2^{13.3}$ distinct sequences), and the owner is detected in 0.96 of such subsets on average.

\paragraph{False claims.}
A claimant with a random key passes the 199-key rule in 0.77\% of tests at 134 games and 0.54\% at 40 (188 programs, 10,000 keys), and two independent random claimants both pass on the same program with probability about $8 \times 10^{-5}$; against 100,000 keys such a pass is matched by a median of 437 other keys, the owner's by at most 8.
Independent use of the same public data (exploratory): on the held-out episodes, each of the 76 owners' keys is tested on the 8 frequency-ordered harnesses (608 comparisons) and on the other owners' keyed seeds with a different key (5,660 comparisons of 76 seeds); the comparisons share programs and keys, so they are counts of tests, not of independent harnesses.
At the 199-key rule 0 and 35 comparisons are flagged (22 owners at least once), at $p \le 0.01$ against 100,000 keys 47, and at $p \le 10^{-3}$ or $10^{-4}$ none (smallest $p$ $1.2 \times 10^{-3}$), while the 114 detected copies among the 132 with $J \ge 1$ stay detected at all three thresholds.
Negative controls: the order-table key flags 0 of 185 tests of runs that never carried it, KGW 0 of 160 templates, the keyed prior 1 of 1,400 tests of programs that never carried it, and another owner's published key 5 of 1,243 (two keys that order most buckets alike).

\paragraph{Three readings of the key} (Table~\ref{tab:endpoints3}, exploratory).
The key is read under the table's published name (the static key endpoint), by content (the largest share of the published same-bucket pairs kept by any dictionary of at least five rows whose keys are object types, whatever its name or encoding, or by the table under its published name) and by detection on the held-out games.
Both static readings are binary at zero: the key is lost when no published same-bucket pair keeps its order (a pruned table whose remaining pairs stay in order keeps it, whatever the share), the reading under the name also requires the table to be read by some code, and the reading by content requires only that the file parses; the share kept is reported as a continuous quantity where given, and the plagiarist-request table of Appendix~\ref{app:plagiarist} uses a half threshold, stated there.
Over the 132 evolved copies with $J \ge 1$ that carried the keyed prior, the name and content readings differ in 9: six tables renamed under the laundering request, two tables whose rows stay in the file while no code reads them, and one index table cut to 0.08 of its pairs.
No copy whose key is gone by content is detected; of the 120 that keep some of it, 6 are not: the two unread tables, a \code{find} that reads rows as sets, one that searches on after the first find, a table \luna pruned ($p = 0.020$) and the index table cut to 0.08.

\begin{table}[htbp]
\centering
\footnotesize
\setlength{\tabcolsep}{3pt}
\begin{tabular}{@{}lcc@{}}
\toprule
& Copies & Laundered \\
\midrule
Programs & 132 & 96 \\
Name and content agree & 123 & 59 \\
Content and detection agree & 126 & 65 \\
Gone by content, detected & 0/12 & 0/29 \\
Kept by content, detected & 114/120 & 36/67 \\
\quad at least half kept, detected & -- & 34/41 \\
\bottomrule
\end{tabular}
\caption{The key under its published name, by content and by detection on the 134 held-out games. Laundered files: the 96 one-shot laundered files played once.}
\label{tab:endpoints3}
\end{table}

\paragraph{A verifier who knows only the agent's visits.}
A customer may not know which receptacles a room holds.
Recomputing detection with the room's kinds taken only from the \code{go to} commands of the episode (prediction registered before the recomputation), the median number of key-dependent searches falls from 80 to 69 in copies and from 79 to 68 in seeds; 109 of the 111 copies with $J \ge 1$ that keep the key under its name are detected (108 with the full room), all 108 seeds, 1 of the 12 copies whose key is gone by content ($p$ from 0.31 to 0.010), and 1 of 432 tests of programs that never carried the key (full room: 1); laundered files give the same decisions as with the full room.
The registered prediction (at least 80\% of copies keeping the key detected, at most 5\% of those that lost it) held for the first part and failed for the second: under the table's name 7 of 21 copies that lost the key are detected (six of them the renamed tables of the laundering run, detected with the full room too), and by content 1 of 12.

\paragraph{Another agent.}
With gpt-oss-120b replacing \qwen as the agent on the held-out games (templates and tool files unchanged), the same 47 of the 50 GEPA copies with $J \ge 1$ under \qwen, \llama and \luna are detected, every detected one at $p = 0.005$, and all 29 seeds; a copy yields 39--84 key-dependent searches (median 70; \qwen: 80), and gpt-oss succeeds in 0.70 of games against \qwen's 0.87--0.88.

\FloatBarrier
\section{Proposer assays}
\label{app:assays}

This appendix supports \S\ref{sec:proposer} and the proposer columns of Table~\ref{tab:proposers}.
Every assay re-runs the evolver's own proposal code on stored feedback with the parent and prompt given, and classifies every rewrite that changes the tool file.
Unless stated, GEPA's joint reflection prompt is used on the 55 stored feedback items of tool-file iterations of the joint-evolution runs whose parent was the seed, 3 samples per item; 95\% intervals resample feedback items; a paired difference is the mean over items with completed rewrites in both conditions of the per-item difference in loss rate, tested by 20,000 sign flips over items.

\paragraph{Encoding and whole-file rewrites.}
With \qwen, the prior table is lost in 16.1\% of rewrites written with names (118 rewrites) and 52.5\% written as indices (120); the identical \code{INSTANCE\_ORDER} in the same files in 24.6\% and 47.5\%; the file breaks in 9.3\% and 10.0\%.
A rewrite is whole-file when the code outside the module-level tables, normalized by Python's \code{ast.unparse}, has a \code{difflib} line ratio below 0.5 against the parent's; \qwen rewrites the named package from scratch in 21\% of rewrites, the index package in 42\% and the neutral package in 45\%, and whole-file rewrites drop the named table in 5 of 22 and the index table in 35 of 45, local edits in 3 of 85 and 16 of 63.
Under \qwen's own model, as the summed surprisal of the table value's tokens with the whole file as prompt, the index table takes 1,753 bits and the named table 1,700.

\begin{table*}[t]
\centering
\footnotesize
\setlength{\tabcolsep}{4pt}
\begin{tabular}{@{}lcccccc@{}}
\toprule
Parent & \multicolumn{3}{c}{\qwen} & \llama & \gptoss & \luna \\
\cmidrule(lr){2-4}
& prior table lost & \code{INSTANCE\_ORDER} lost & key kept & not verbatim & not verbatim & lost \\
\midrule
names & 16.1\% & 24.6\% & 0.864 & 9.7\% & 100\% & 50.9\% \\
names + provenance comment & 18.9\% & 25.2\% & 0.834 & 12.7\% & 98.8\% & -- \\
names + unrelated comment & 20.3\% & 34.1\% & 0.833 & 15.2\% & 99.4\% & -- \\
names + keep instruction & 9.4\% & 11.1\% & 0.915 & 8.5\% & 90.3\% & -- \\
indices & 52.5\% & 47.5\% & 0.525 & 6.7\% & 99.4\% & 62.7\% \\
indices + provenance comment & 23.3\% & 25.2\% & 0.805 & 6.1\% & 100\% & 61.8\% \\
indices + unrelated comment & 54.8\% & 53.2\% & 0.484 & 6.1\% & 98.8\% & 64.5\% \\
indices + keep instruction & 11.4\% & 8.6\% & 0.923 & 7.3\% & 97.0\% & 38.2\% \\
\bottomrule
\end{tabular}
\caption{Comments of the same length placed above the prior table. \qwen, \llama and \gptoss under GEPA's joint prompt, \luna under the tool-only prompt (2 samples per item). ``Key kept'': share of the published same-bucket pairs whose order survives, averaged over rewrites.}
\label{tab:comments}
\end{table*}

\paragraph{Comments} (Table~\ref{tab:comments}).
For \qwen's index table, the unrelated comment has no detectable effect ($+6.5$ points against no comment) and the keep instruction cuts deletion to a fifth ($-43.7$ points; $-12.9$ against the provenance comment, $p = 0.03$, $p_{\mathrm{H}} = 0.40$); the provenance comment also protects the identical \code{INSTANCE\_ORDER} in the same file (47.5\% to 25.2\%).
Against \qwen with the same prompt and items, \luna loses the named table 39.4 points and the index table 23.1 points more often.
\llama's 1,485 rewrites never lose \code{INSTANCE\_ORDER} or the tool file, every changed table is a row edit, and no comment or prompt changes it detectably ($p \ge 0.24$).

\paragraph{Proposers.}
gemma changes the prior table in 100\% and 46.7\% of rewrites (names, indices) by editing about 6 of 61 rows, most of them copying errors, and keeps 0.983 and 0.987 of the key.
\luna (all 550 calls return a parseable new file) loses the named table by editing 40 and deleting 16, the index table by deleting 60 and editing 9; its edits drop 1,593 rows from 40 named tables while the rows it keeps stay in the owner's order (3,705 of 3,893 same-bucket pairs).
\gptoss keeps 0.002--0.008 of the key and breaks the file in 0.6--6\% of rewrites; none of its 1,815 replies reached the token limit, and its new files have a median of 6,400--6,600 characters against 11,000 and 14,500 for the parents, so its elision does not follow from the output budget.
Its elided files parse (exploratory): of 312 joint-prompt and 327 tool-only rewrites that are not broken, \code{LIKELY\_PLACES} becomes an empty dictionary (\code{\{ \# ...unchanged... \}}) in 222 and 236, so \code{find} falls back to the default order, a set \code{\{ ... \}} in 65 and 68, on which \code{.get} fails at every \code{find} call, a dictionary with fewer rows in 22 and 21, and no assignment in 4.
Among the deployed \gptoss copies, one holds the set and its \code{find} failed in all 182 calls on held-out games, after which the agent searched with \code{act}; another guards against it with a type check.
A reminder to output the complete file brings its loss to 95.8\% and 96.4\% (key kept 0.058, 0.042), the 15-row table to 66.1\% (0.383), both together to 43.6\% (0.601).
With ShinkaEvolve's diff patches on the same 110 tasks (3 samples), 319 of 330 \gptoss patches applied; they lose the prior table in 4.3\% (named) and 3.2\% (index) of rewrites, nearly all through broken files, and keep 0.96 and 0.97 of the key; the registered prediction (at most 30\% lost, key kept at least 0.7) held.

\paragraph{Prompt, feedback and history.}
Under GEPA's tool-only prompt, \qwen loses the named and index prior tables in 10.1\% (4.3--16.7) and 37.0\% (27.3--46.7) of rewrites and \code{INSTANCE\_ORDER} in 25--46\%; key kept 0.929 and 0.684.
Given the stored ten-game feedback of the minibatch-10 runs whole or reduced to three games, \qwen loses the prior table in 27.7\% against 27.8\%.
On 92 parents from ShinkaEvolve programs after 0, 1--2, 3--4 or 5 or more earlier tool-file edits (logistic regression clustered by run), the named table is lost at 0.40, 0.13, 0.33 and 0.19 by bin and the index table at 0.25, 0.39, 0.55 and 0.50 at 24,000 tokens; at 48,000, 0.08, 0.17, 0.38, 0.26 and 0.58, 0.47, 0.44, 0.41.

\paragraph{Edit format.}
ShinkaEvolve's prompt construction with one patch type, its LLM client and its patch application on the same items and parents: diff patches lose the prior table in 0.7\% and 6.7\% (names, indices) and \code{INSTANCE\_ORDER} in 2.7\%, full rewrites 45.8\% and 78.9\%, and 70.4\% and 76.8\%.
Deleting proposals call the table hard-coded or static in 60--83\% of their stated reasons and, in full rewrites, propose discovering receptacles from the room in 82--86\%; keeping proposals call it hard-coded in 4--53\%.

\paragraph{Cut-off replies.}
Only \qwen reached the 24,000-token limit, in 12--42\% of replies per assay and package (27\% in the history assay); counting every cut-off reply as keeping or as losing the table bounds each loss rate (tool-only prompt: names 0.061--0.461, indices 0.242--0.588), and only the laundering contrast is separated by both bounds.
At 48,000 tokens 2.9--6.1\% are cut off; the tool-only prompt loses the named and index tables in 5.7\% and 28.0\%, and under the joint prompt the index table with the unrelated comment, the keep instruction and the two provenance comments in 41.8\%, 16.1\%, 18.2\% and 20.1\%.

\paragraph{License notice and safety check.}
The named package was given a five-line header (a copyright line for a fictitious project, an Apache-2.0 notice asking derived versions to keep it, and an attribution line) and a function \code{\_checked\_step} that runs a command only if the environment admits it, through which \code{find} and \code{place} issue every command; the tools issue only admitted commands, so the check never fires and the file behaves as before.
Under the tool-only prompt (Table~\ref{tab:safety}), \qwen kept the notice verbatim in 98\% of rewrites and the check in 88\%; \luna rewrote the notice into the standard Apache wording in every rewrite but kept the copyright and attribution lines in 93\% and the check in 91\%; \gptoss kept the notice verbatim in 89\% and the check in 98\% while eliding the table in every rewrite.
Of the registered predictions, \luna changed the notice more often than \qwen (paired difference 0.98, $p < 0.0001$), and the other three failed: \qwen changed the notice in 2\% (predicted at least 10\%) and dropped the check in 12\% (at least 20\%), and \gptoss dropped the check in 2\% (at least 50\%).
With the notice and check in the file, \qwen lost the prior table in 3.7\% of rewrites, against 10.1\% for the same package without them$^\dagger$.

\begin{table}[htbp]
\centering
\footnotesize
\setlength{\tabcolsep}{2.5pt}
\begin{tabular}{@{}lccc@{}}
\toprule
& \qwen & \luna & \gptoss \\
\midrule
Rewrites & 107 & 110 & 165 \\
Prior table lost & 3.7\% & 85\% & 100\% \\
Notice verbatim & 98\% & 0\% & 89\% \\
Copyright, attribution kept & 99--100\% & 93\% & 93--99\% \\
Check kept & 88\% & 91\% & 98\% \\
\bottomrule
\end{tabular}
\caption{License notice and command check in the named tool file, selection off, tool-only prompt.}
\label{tab:safety}
\end{table}

\FloatBarrier
\section{Selection}
\label{app:selection}

This appendix supports \S\ref{sec:selection}.
A carrier is intact when its literal is unchanged and some code still reads it; a literal that stays but is no longer read counts as lost (5 rejected proposals and no pool candidate in the joint-evolution runs).
Table~\ref{tab:accept} gives the acceptance of tool-file proposals whose parent missed at least one minibatch game at 3, 10 and 30 games per decision; $p$ comes from Cochran--Mantel--Haenszel tests stratified by root, and because proposals of a root share parents and games we also resample whole roots (20,000 samples).

\begin{table}[htbp]
\centering
\scriptsize
\setlength{\tabcolsep}{2pt}
\begin{tabular}{@{}llccc@{}}
\toprule
$n$ & Carrier & Accepted, & Root-resampled & $p$ \\
& & lost\,/\,kept & difference & \\
\midrule
3 & prior table & 8/20\,/\,32/76 & $-0.02$ ($-0.13$, 0.17) & 0.85 \\
3 & tool file & 4/9\,/\,36/87 & $+0.03$ ($-0.27$, 0.33) & 0.89 \\
3 & order table & 12/23\,/\,25/66 & $+0.14$ ($-0.09$, 0.34) & 0.60 \\
10 & prior table & 1/20\,/\,28/64 & $-0.39$ ($-0.52$, $-0.27$) & 0.002 \\
10 & tool file & 0/12\,/\,29/74 & $-0.39$ ($-0.49$, $-0.28$) & 0.023 \\
10 & order table & 2/19\,/\,27/65 & $-0.31$ ($-0.48$, $-0.09$) & 0.011 \\
30 & prior, names & 4/17\,/\,60/192 & -- & -- \\
30 & prior, indices & 13/53\,/\,59/176 & -- & -- \\
30 & neutral table & 10/32\,/\,14/60 & -- & -- \\
\bottomrule
\end{tabular}
\caption{Acceptance of tool-file proposals with a beatable parent at $n$ games per decision (order table: \code{INSTANCE\_ORDER}): joint-evolution runs of the named package at 3 and 10 games (8 roots each), prospective tool-only runs at 30 games (8 roots each). At 30 games accepted edits lost the named table in 4 of 64, the index table in 13 of 72 and the neutral table in 10 of 24.}
\label{tab:accept}
\end{table}

\paragraph{What selection passes on.}
Share of tool-file rewrites of an intact parent that lose the carrier, among all proposals and among accepted ones, at minibatch 3: prior table 49 of 318 (0.154) and 8 of 40 (0.200); index package 57 of 168 and 5 of 20; \code{INSTANCE\_ORDER} 76 of 291 and 12 of 37; neutral package 85 of 145 and 8 of 14.
At minibatch 10: prior table 31 of 123 (0.252) and 1 of 29 (0.034); whole tool file 19 of 125 and 0 of 29.
The per-proposal hazard of the prior table is 2.5\% (8 of 318) at minibatch 3 and 0.8\% (1 of 123) at 10; no pairwise difference in hazard between templates, minibatches or encodings reaches 0.05 when whole roots are permuted.

\paragraph{Fresh-game model.}
The 98 logged proposals (all 49 that deleted the prior table from an intact parent in the minibatch-3 roots, and 49 random keeping ones) were each played once with their parents on games 80--119.
A minibatch is $n$ of these games drawn without replacement and shared by parent and child, and the child is accepted when it wins strictly more of them; predicted acceptance of deleting and keeping proposals with a beatable parent is 0.27 and 0.38 at $n = 3$, 0.19 and 0.30 at $n = 10$, and 0.10 and 0.22 at $n = 30$.
Deleting proposals change success by $-0.111$ on average and keeping ones by $-0.060$.
Played on GEPA's own 60 training games, deleting proposals are accepted 0.81 (0.67--0.97), 0.66 (0.50--0.84) and 0.51 (0.30--0.78) times as often as keeping ones at 3, 10 and 30 games.

\paragraph{The final pick.}
Drawing $m$ of a run's 40 validation games (one subset for the whole pool, 2,000 draws; ties go to the earlier candidate, as in GEPA's \code{best\_idx}), the probability that the highest-mean candidate carries the prior table rises from 0.954 at $m = 5$ to 1.000 at $m = 40$ against a pool share of 0.823 in the joint-evolution runs; for \code{INSTANCE\_ORDER} it goes from 0.917 to 0.875 against 0.747.
In the pools of the 30-game runs, 60 of 73 named and 59 of 77 index candidates with $J \ge 1$ keep the table (validation 0.935 and 0.929 against 0.925 and 0.874), and all 13 deployed copies with $J \ge 1$ do$^\dagger$.
Across the joint-evolution runs, 358 proposals start from a parent that has already lost a table and one restores it (a broken file repaired); GEPA's reflection prompt shows only the current candidate, so a deleted carrier has no route back.

\FloatBarrier
\section{Prospective tests and prediction ledger}
\label{app:prospective}

This appendix supports \S\ref{sec:prospective} and the marks $^\dagger$ in the body.

\paragraph{Frozen parameters.}
Before any run of a condition started, $d$, $\rho$ and the predictions were written into files whose SHA-256 hashes were recorded in the project log (supplementary material).
$d$ comes from the selection-off assays with the tool-only prompt, 3 samples per item, plus the history slope of the named and index tables at 24,000 tokens.
$\rho$ by way of loss comes from the 98 proposals re-run on fresh games (Appendix~\ref{app:selection}; relative to keeping proposals: broken file 0.467, deleted table 0.771, edited rows 0.984 at 3 games; 0.105, 0.397 and 0.998 at 30 games; AgentMark's sampling code 1.43), with the probability that the parent is beatable (0.302 and 0.968 at 3 and 30 games).
A variant with $\rho$ from GEPA's training games was frozen before any \qwen result was read; it falls between A and B on every summary and is listed in the supplement.
The second test (longer runs, laundering inside GEPA, shifted tasks) reuses these counts, slopes and $\rho$ without refitting, and predicts the key endpoint: A$'$ is the share of frozen assay rows that erase the key, B$'$ weights erasing ways of loss by the frozen $\rho$ and key-keeping row edits by $\rho$ for row edits.

\begin{table*}[t]
\centering
\scriptsize
\setlength{\tabcolsep}{3pt}
\begin{tabular}{@{}lllllll@{}}
\toprule
Condition & Carrier & Proposals & Accepted edits losing it & $q_A$\,/\,$q_B$ & Key lost: obs., A$'$, B$'$ & Inside \\
\midrule
\qwen, names, 3 g. & prior table & 46/568 & 9/68 = 0.13 (0.05--0.23) & 0.101 / 0.083 & 0.074, 0.091, 0.073 & A, B \\
& order table & 102/568 & 11/67 = 0.16 (0.10--0.25) & 0.253 / 0.196 & & B \\
\qwen, indices, 3 g. & prior table & 191/582 & 20/68 = 0.29 (0.18--0.42) & 0.370 / 0.346 & 0.265, 0.333, 0.305 & A, B \\
& order table & 177/590 & 18/67 = 0.27 (0.17--0.37) & 0.333 / 0.302 & & A, B \\
\qwen, neutral, 3 g. & order table & 219/552 & 27/90 = 0.30 (0.21--0.41) & 0.455 / 0.392 & & B \\
\qwen, AgentMark, 3 g. & sampling code & 67/309 & 12/56 = 0.21 (0.12--0.38) & 0.345 / 0.430 & & A \\
& order table & 101/253 & 17/43 = 0.40 (0.22--0.61) & 0.455 / 0.392 & & A, B \\
\qwen, names, 30 g. & prior table & 17/211 & 4/64 = 0.06 (0.02--0.13) & 0.101 / 0.058 & 0, 0.091, 0.047 & A, B \\
& order table & 30/213 & 9/65 = 0.14 (0.05--0.23) & 0.253 / 0.124 & & B \\
\qwen, indices, 30 g. & prior table & 54/233 & 13/72 = 0.18 (0.12--0.26) & 0.370 / 0.279 & 0.139, 0.333, 0.227 & none \\
& order table & 42/233 & 13/72 = 0.18 (0.12--0.26) & 0.333 / 0.225 & & B \\
\qwen, neutral, 30 g. & order table & 32/92 & 10/24 = 0.42 (0.33--0.59) & 0.455 / 0.249 & & A \\
\midrule
\luna, names, 3 g. & prior table & 135/169 & 19/24 = 0.79 (0.64--0.86) & 0.509 / 0.489 & 0.25, 0.146, 0.117 & none \\
\luna, indices, 3 g. & prior table & 257/345 & 39/48 = 0.81 (0.73--0.89) & 0.627 / 0.574 & 0.75, 0.573, 0.512 & none \\
\llama, names, 3 g. & prior table & 63/288 & 7/45 = 0.16 (0.02--0.34) & 0.121 / 0.120 & 0, 0, 0 & A, B \\
\gptoss, names, 3 g. & both tables & 207/207 & 22/22 = 1.0 & 1.0 / 1.0 & 0.955, 0.946, 0.944 & A, B \\
\midrule
Longer runs, names & prior table & -- & 7/123 = 0.057 (0.031--0.091) & -- & 0.049, 0.089, 0.072 & \\
Longer runs, indices & prior table & -- & 17/105 = 0.162 (0.124--0.222) & -- & 0.152, 0.424, 0.368 & \\
Shifted, names & prior table & -- & 8/53 = 0.15 (0.06--0.24) & -- & 0.06 & \\
Shifted, indices & prior table & -- & 12/54 = 0.22 (0.15--0.33) & -- & 0.20 & \\
\bottomrule
\end{tabular}
\caption{Literal and key endpoints of the prospective tests by condition and carrier (3 g., 30 g.: games per decision; order table: \code{INSTANCE\_ORDER}). ``Proposals'': tool-file rewrites of a parent whose carrier was intact that lose it; ``accepted edits'': accepted tool edits, intervals resampling roots; $q$ is the mean frozen prediction over accepted edits; ``inside'': models whose literal prediction lies inside the interval. \llama never lost \code{INSTANCE\_ORDER} or the tool file in 299 tool edits; the tool file broke in 3 of 594 accepted \qwen edits.}
\label{tab:prospective-rows}
\end{table*}

\paragraph{Lineages} (Table~\ref{tab:brier}).
Each deployed lineage and carrier gives one observation, kept or lost, against the frozen $S(J)$ of A and B; the constant C predicts, for each lineage, the share of the other lineages in its group that kept the carrier (exploratory, and fitted to the outcomes, so it is a reference scale and no prediction).
B beats or ties A in every group; it does not beat C on the first test, and C beats both on the longer runs, where B predicted index losses that GEPA's final pick did not deploy: accepted edits there lost the index key under its name at 0.15 against B's 0.37, and by content at 0.08.
Binned by B's prediction (\qwen, both tests, literal endpoint), lineages predicted at 0--0.2 kept the carrier in 5 of 7 cases, at 0.4--0.6 in 0.83 and at 0.8--1 in 0.95 (mean prediction 0.95): B is calibrated where it predicts survival and too pessimistic for the index table.

\begin{table}[htbp]
\centering
\footnotesize
\setlength{\tabcolsep}{2.4pt}
\begin{tabular}{@{}lccccc@{}}
\toprule
Group & Pairs & A & B & C & C $-$ B \\
\midrule
\qwen, first test & 112 & 0.151 & 0.126 & 0.137 & $+0.011$ \\
\quad $J \ge 1$ & 76 & 0.222 & 0.186 & 0.186 & $0.000$ \\
\quad key endpoint & 124 & 0.139 & 0.113 & 0.138 & $+0.025$ \\
\qwen, longer runs & 24 & 0.315 & 0.282 & 0.083 & $-0.199$ \\
\luna & 24 & 0.159 & 0.142 & 0.242 & $+0.100$ \\
\llama & 12 & 0.021 & 0.020 & 0.000 & $-0.020$ \\
\gptoss & 12 & 0.000 & 0.000 & 0.297 & $+0.297$ \\
\bottomrule
\end{tabular}
\caption{Brier score of deployed lineage--carrier pairs (literal endpoint unless stated) under the frozen Model A and B and a leave-one-out constant C. Intervals resampling roots: C $-$ B $-0.030$ to 0.054 (\qwen first test), $-0.351$ to $-0.050$ (longer runs); A $-$ B is positive or zero in every row (\qwen first test 0.025, 0.010 to 0.040).}
\label{tab:brier}
\end{table}

\begin{figure}[htbp]
\centering
\includegraphics[width=\columnwidth]{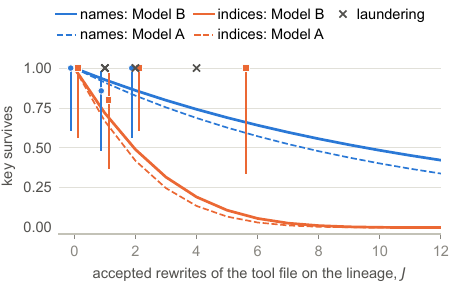}
\caption{Key survival on deployed \qwen lineages (3 games per decision) after $J$ accepted tool-file rewrites: frozen predictions B (solid) and A (dashed); prospective and longer-run lineages binned by $J$ (key endpoint, Wilson intervals); crosses: laundering in GEPA (renamed tables, key by content).}
\label{fig:survival}
\end{figure}

\paragraph{Selection apart from proposer rates (exploratory).}
Among \qwen's proposals whose parent could be beaten, deleting the prior table was accepted 0.81 (0.50--1.16) times as often as keeping it at 3 games (frozen 0.77), edited rows 1.13 (0.48--1.96; frozen 0.98) and broken files in 0 of 4 cases; at 30 games deletions were accepted 0.65 (0.31--1.17) times as often (frozen 0.40), edited rows 1.46 (0.96--2.15) and key-erasing rewrites 0.54 (0.29--0.90), none of 10 with the named table.
On single accepted edits the frozen B's log-likelihood is above A's by 9.1 ($-1.4$ to 19.0) on the literal endpoint and by 18.3 (7.9 to 28.5) on the key endpoint$^\dagger$.
Recomputing both versions with the runs' own $d$ ($A_{\mathrm{run}}$: $d$; $B_{\mathrm{run}}$: the same $d$ weighted by the frozen $\rho$), $B_{\mathrm{run}} - A_{\mathrm{run}}$ in log-likelihood on accepted tool edits is $-5.93$ ($-14.65$ to 1.89) on the literal endpoint and $-2.06$ ($-10.97$ to 6.55) on the key endpoint, while $A_{\mathrm{run}} - A$ is $+19.96$ (9.30 to 34.09): the frozen B gained on single edits mostly by offsetting the overestimated frozen $d$.
On deployed lineages the selection term still lowers the Brier score by 0.013 (0.001 to 0.024), carried by the 30-game conditions.
The proposer loses the named table more often when the parent missed a minibatch game: 0.130 against 0.057 ($+0.072$, 0.031 to 0.123), the same as the loss per accepted edit (9 of 68).
\luna's proposals from the seed itself lose the table in 0.80 and 0.69 of the runs' rewrites against 0.51 and 0.63 in the assay with the same parents and prompt but \qwen's feedback.

\paragraph{The second test.}
Longer runs (Figure~\ref{fig:survival}): the pools hold 16--35 candidates and lineages up to $J = 7$, but the deployed candidate (validation 0.925--1.000) has $J$ = 1, 1, 1, 2, 2, 2 (names) and 1, 2, 2, 2, 5, 6 (indices); 5 of 6 named and 6 of 6 index copies keep the key (B 5.4 and 2.4; 6 of 6 has probability 0.0005 under B); the named copy without the key still holds the table's rows but no code reads them, and it is the one copy not detected ($p = 0.975$).
With the 48,000-token history slopes in place of the frozen ones, B predicts 5.3 and 2.8 copies with the key$^\dagger$.
Laundering inside GEPA: all 20 accepted tool edits from a parent holding \code{LIKELY\_PLACES} removed it by name and none erased the key by content, so the frozen prediction held by name (20 of 20) and failed by content; all 8 copies are detected ($p = 0.005$, 75--92 key-dependent searches).
Shifted tasks: 308 training games are \code{look\_at\_obj\_in\_light} games; training and validation games were drawn from them, $\rho = 1$ was frozen for every way of loss, and the index copies that lost the key (2 of 6) are not detected; held-out success changed by $+0.016$ and $-0.006$.

\paragraph{Prediction ledger.}
\label{app:tests}
Table~\ref{tab:ledger} lists every prediction registered for the prospective tests and the later experiments, and Table~\ref{tab:tests} every hypothesis test with $p_{\mathrm{H}}$.
Predictions were written into the project log on 28 September (selection), 29 September (proposer assays), 1--2 October (comment, feedback and history assays, prospective test, ShinkaEvolve replication, telecom runs), 5 October (placement, laundering, agent swap, cooking), 6 October (second prospective test, laundered files, telecom bound, comparison without a key) and 7 October (license and safety check, plagiarist objectives, fewer games, diff patches for \gptoss, partial layout); the ShinkaEvolve edit-format assay, the provenance comment and the \luna and \gptoss assays had none.
The log and frozen files carry no third-party timestamp.
Fourteen of the 31 tests stay below 0.05 after correction; among those that do not, the minibatch-10 filtering of broken files has a root-cluster interval that excludes zero, and the cost of deleting proposals agrees with the table's separately measured value (0.051 against 0.046).

\begin{table*}[p]
\centering
\scriptsize
\setlength{\tabcolsep}{3pt}
\begin{tabularx}{\textwidth}{@{}p{2.6cm} Y Y c@{}}
\toprule
Experiment & Prediction written before the data were read & Observed & Held \\
\midrule
Comment assay & \qwen: unrelated comment within 10 points of none; keep instruction below none and above the provenance comment & 54.8\% against 52.5\%; 11.4\%, below 23.3\% & partly \\
Feedback size & more loss and breakage with ten games of feedback than three & 27.7\% against 27.8\% & no \\
History assay & loss does not fall with earlier edits (slope $\ge 0$) & index $+0.165$, named $-0.011$ & partly \\
Tool-only prompt & loss within 10 points of the joint prompt & $-9.6$ points & yes \\
AgentMark assay & sampling code removed in at least 5\% of rewrites & 1.8\% (32.7\% unparseable) & no \\
Prospective test & accepted-edit loss inside the root interval for A and B; B's log-likelihood above A's & B inside in 9 of 12 \qwen rows, A in 7; $+9.1$ ($-1.4$ to 19.0) & partly \\
\quad other proposers & observed loss inside the interval & \llama and \gptoss yes, \luna no & partly \\
ShinkaEvolve replication & index table lost about 1.9 times as often, root permutation $p < 0.05$ & 2.6 times, $p = 0.003$ ($p_{\mathrm{H}} = 0.051$) & yes \\
& diff patches below 5\%; full rewrites about 25\% and 42\% & diff 0.5\% and 7.6\%; full 21.9\% and 38.1\% & partly \\
& programs chosen as parents lose the named table less often & 0.056 against 0.078, $p = 0.51$ & no \\
& deployed programs no better than their seeds on validation & 18 of 20 roots & yes \\
AgentMark in one file & no unloadable deployed file; under 2\% fenced rewrites; accepted edits lose the code in 0.10--0.35; at least 7 of 12 detected & 0; 0/734; 0.05; 7 of 12 & partly \\
Telecom & checklist raises success; detectable at publication & unchanged; $p = 0.005$ & partly \\
& index plan lost 1.5 times as often; accepted edits at the proposer's rate; breakage under 5\%; intact copies detected, changed ones not & 0.93 against 0.94; 1.00 against 0.94; 0; yes & partly \\
\midrule
Laundering, \qwen & prior table lost in 20--40\% (names) and 50--70\% (indices); key kept below 0.85 (names) & name gone in 95--98\%, key kept 0.89 and 0.52 by content & partly \\
Laundering, \llama & more changes, key kept at least 0.95 & loss 0.17 and 0.05 ($p \ge 0.18$); key 0.96 & partly \\
Laundering, \gptoss & loss still near total & 100\% and 98.8\% & yes \\
Laundering, all & less code kept than without the sentence & lower for three proposers, unchanged for \llama & partly \\
Agent swap & copies detected under \qwen and with at least 20 key-dependent searches under gpt-oss are all detected; detection agrees with the key in at least 90\% & 47 of 47; 49 of 50 & yes \\
KGW docstring & KGW detected in at most half of the exposed copies, the key in at least 90\% & 1 of 7; 7 of 7 & yes \\
Data file & accepted edits lose it in at most 0.10; at least 90\% detected & 6 of 51; 6 of 7 & no \\
Comments in evolution & accepted edits lose the index table in at most 0.20 (provenance) and 0.15 (keep); 7 of 8 detected & 0.12 and 0; all detected & yes \\
Cooking gate & table worth less than 0.03 & $-0.033$ & yes \\
\midrule
Laundered files played & \qwen, \luna, \gptoss lose 0 to 0.10, \llama $-0.02$ to $+0.02$; at least half lose under 0.06 & $-0.060$, \llama $-0.026$; 44 of 72 & partly \\
& no file with the key erased detected; files keeping it detected in 90\% & 13 renamed tables detected; elided tables 0 of 6 & no \\
Laundering in GEPA & accepted edits erase the key in at least 0.80; at most 25\% detected; success change $-0.08$ to $+0.02$ & 20 of 20 by name, 0 of 20 by content; 6 of 6; $-0.001$ & partly \\
Longer runs & median deployed $J$ at least 4; copies keeping the key inside B's interval, B above A; detection agrees with the key in 90\% & median 2; names yes, indices no (6 of 6, probability 0.0005); 12 of 12 & partly \\
48,000 tokens & cut-off below 10\%; index lost twice as often (tool-only); keep and provenance protect; index slope $> 0$ & 2.9--6.1\%; 28.0 against 5.7\%; $p \le 0.0002$; $-0.12$ & partly \\
Shifted tasks & more literal loss than on the original tasks; copies keeping the key inside the $\rho = 1$ interval & 8/53 and 12/54; indices yes, names 6 of 6 (probability 0.18) & partly \\
Telecom order bound & best order saves under 0.5 checks per task & 1.35 & no \\
Comparison without a key & same-data harnesses as similar as copies (AUC at most 0.6) & 0.81 & no \\
\midrule
License notice, check & \qwen changes or drops the notice in at least 10\%; \luna more often than \qwen; \qwen drops the check in at least 20\%, \gptoss in at least 50\% & 2\%; 100\% against 2\%; 12\%; 2\% & partly \\
Plagiarist requests & clean-up: \qwen loses the table by content in at least 70\%; more general: more loss than without a request; \luna keeps less than 0.3 of the key under both & 94\% and 100\%; $+0.69$ and $+0.65$, $p_{\mathrm{H}} \le 0.003$; 0.00 & yes \\
Fewer games & success rises with the games; 35-game table below 0.86; 700-game table within 0.02 of 0.889 & 0.861, 0.879, 0.889 & partly \\
\gptoss, diff patches & prior table lost in at most 30\% (names); key kept at least 0.7 & 4.3\%; 0.96 & yes \\
Partial layout & copies keeping the key detected in at least 80\%; those that lost it in at most 5\% & 109 of 111; 7 of 21 by name, 1 of 12 by content & partly \\
\bottomrule
\end{tabularx}
\caption{Prediction ledger: registered predictions and outcomes, in the order they were written (1 October to 7 October); earlier predictions appear with their tests in Table~\ref{tab:tests}.}
\label{tab:ledger}
\end{table*}

\begin{table*}[t]
\centering
\scriptsize
\setlength{\tabcolsep}{3pt}
\begin{tabular}{@{}lllcc@{}}
\toprule
Test & Comparison & Prediction & $p$ & $p_{\mathrm{H}}$ \\
\midrule
Broken file, minibatch 3 & accepted 4/9 against 36/87 (CMH by root) & 0.67 times as often & 0.89 & 1.0 \\
Broken file, minibatch 10 & accepted 0/12 against 29/74 (CMH by root) & 0.40 times as often & 0.023 & 0.30 \\
Prior table, minibatch 10 & accepted 1/20 against 28/64 (CMH by root) & 0.73 times as often & 0.002 & 0.036 \\
Per-proposal hazard & minibatch 3 against 10 (root permutation) & slight rise at 10 & 0.097 & 0.87 \\
Cost of deleting proposals & fresh games, labels permuted within roots & about the table's value & 0.036 & 0.40 \\
Neutral table, minibatch 10 & accepted 2/19 against 27/65 (CMH by root) & none & 0.011 & 0.17 \\
Prompt: broken file & tool-only against joint prompt (sign flip) & none & $<$0.0001 & $\le$0.003 \\
Prompt: prior table & tool-only against joint prompt (sign flip) & within 10 points & 0.035 & 0.40 \\
Feedback size & three against ten trajectories (sign flip) & more loss with ten & 0.99 & 1.0 \\
History, index table & slope per earlier edit (clustered by run) & slope $\ge 0$ & 0.004 & 0.059 \\
History, named table & slope per earlier edit (clustered by run) & slope $\ge 0$ & 0.94 & 1.0 \\
Encoding, ShinkaEvolve diff & names against indices (paired) & none & 0.011 & 0.17 \\
Encoding, GEPA & names against indices (paired) & more loss with indices & $<$0.0001 & $\le$0.003 \\
Encoding, ShinkaEvolve full & names against indices (paired) & none & $<$0.0001 & $\le$0.003 \\
Provenance comment, \qwen & index table with and without (paired) & none & 0.0007 & 0.014 \\
Unrelated comment, \qwen & index table with and without (paired) & within 10 points & 0.33 & 1.0 \\
Keep instruction, \qwen & index table with and without (paired) & less loss & $<$0.0001 & $\le$0.003 \\
Keep against provenance, \qwen & index table (paired) & more loss with the keep instruction & 0.033 & 0.40 \\
Keep instruction, \luna & index table with and without (paired) & none & 0.0009 & 0.017 \\
Short table, \gptoss & 15 against 61 rows (paired) & none & $<$0.0001 & $\le$0.003 \\
48k: keep instruction & against the unrelated comment, index table (paired) & less loss & $<$0.0001 & $\le$0.003 \\
48k: provenance comment & against the unrelated comment, index table (paired) & less loss & 0.0001 & 0.003 \\
48k: history, index table & slope per earlier edit (clustered by run) & slope $> 0$ & 0.28 & 1.0 \\
48k: history, named table & slope per earlier edit (clustered by run) & none & 0.0004 & 0.008 \\
ShinkaEvolve replication & index against named loss per root (exact permutation) & more loss with indices & 0.003 & 0.051 \\
\quad parent choice & named table, chosen against never chosen (by root) & less loss when chosen & 0.51 & 1.0 \\
Telecom, named checklist & success with and without (McNemar) & none & 1.0 & 1.0 \\
Telecom, index checklist & success with and without (McNemar) & none & 0.26 & 1.0 \\
License notice & \luna against \qwen, notice changed (paired) & more change by \luna & $<$0.0001 & $\le$0.003 \\
More-general request, names & table lost by content, against no request (paired) & more loss & $<$0.0001 & $\le$0.003 \\
More-general request, indices & table lost by content, against no request (paired) & more loss & $<$0.0001 & $\le$0.003 \\
\bottomrule
\end{tabular}
\caption{Hypothesis tests whose $p$ the paper reports, with the prediction written before the data were read (``none'': exploratory, marked $^\dagger$ in the body), and $p_{\mathrm{H}}$ after Holm's correction across all of them. Root-cluster intervals for the minibatch-10 comparisons are in Table~\ref{tab:accept}. CMH: Cochran--Mantel--Haenszel test.}
\label{tab:tests}
\end{table*}

\section{Deployed copies and the held-out test}
\label{app:deployed}
\label{app:heldout}

This appendix supports \S\ref{sec:deployed}.
Each frozen program ran once on the 134 games of \code{eval\_out\_of\_distribution}; differences are paired by game with bootstrap intervals over games, and a copy identical to its seed reuses the seed's games.
Table~\ref{tab:heldout} gives every arm; per-copy listings, $J$ per lineage and detection on validation games are in the supplementary material.
On validation games the keyed prior gives the same decision as on held-out games for 75 of the 76 copies of Table~\ref{tab:consolidated}.

\begin{table*}[t]
\centering
\scriptsize
\setlength{\tabcolsep}{3pt}
\begin{tabular}{@{}lccccc@{}}
\toprule
Arm & Copies ($J \ge 1$) & Seed $\rightarrow$ deployed & Above\,/\,below 0 & Keyed prior & Order table\,/\,other \\
\midrule
\multicolumn{6}{@{}l}{\emph{Joint evolution, ShinkaEvolve and HarnessX}} \\
GEPA, weak template, order table + KGW & 4 & 0.726 $\rightarrow$ 0.879 & 2\,/\,0 & -- & 2/4, KGW 0/4 \\
GEPA, weak template, AgentMark & 4 & 0.431 $\rightarrow$ 0.772 & 4\,/\,0 & -- & AgentMark 3/4 \\
GEPA, weak template, names\,/\,indices & 4 + 4 & 0.754, 0.785 $\rightarrow$ 0.911, 0.957 & 7\,/\,0 & 8/8 & 6/8 \\
GEPA, names, minibatch 3\,/\,10 & 8 + 8 & 0.904 $\rightarrow$ 0.924, 0.934 & 5\,/\,0 & 16/16 & 15/16 \\
GEPA, indices & 4 & 0.901 $\rightarrow$ 0.927 & 2\,/\,0 & 4/4 & 4/4 \\
GEPA, neutral\,/\,frequency order & 4 + 4 & 0.873, 0.907 $\rightarrow$ 0.916, 0.914 & 2\,/\,0 & 0/4 (no key) & 8/8 \\
GEPA, \llama\,/\,\gptoss proposer & 4 + 4 & 0.894 $\rightarrow$ 0.860, 0.912 & 2\,/\,2 & 4/4, 1/4 & 5/8 \\
ShinkaEvolve, names\,/\,indices & 4 + 4 & 0.894, 0.901 $\rightarrow$ 0.892, 0.871 & 0\,/\,1 & 5/8 & 4/8 \\
HarnessX, weak template, KGW\,/\,AgentMark & 4 + 4 & 0.726, 0.431 $\rightarrow$ 0.761, 0.638 & 3\,/\,0 & -- & 4/4, AgentMark 4/4 \\
HarnessX, names & 4 & 0.894 $\rightarrow$ 0.827 & 1\,/\,2 & 4/4 & 4/4 \\
\midrule
\multicolumn{6}{@{}l}{\emph{Prospective test (tool-only GEPA)}} \\
\qwen names\,/\,indices, 3 games & 12 + 12 (6 + 7) & 0.872, 0.871 $\rightarrow$ 0.874, 0.876 & 0\,/\,1 & 23/24 (12/13) & 22/24 \\
\qwen names\,/\,indices, 30 games & 8 + 8 (6 + 7) & 0.875, 0.865 $\rightarrow$ 0.872, 0.874 & 0\,/\,0 & 16/16 (13/13) & 16/16 \\
\qwen neutral, 3\,/\,30 games & 12 + 8 (10 + 8) & 0.852, 0.845 $\rightarrow$ 0.845, 0.868 & 0\,/\,0 & -- & 11/20 \\
\qwen AgentMark, two\,/\,one file & 12 + 12 (10 + 12) & 0.836, 0.830 $\rightarrow$ 0.849, 0.856 & 4\,/\,1 & -- & 20/24, AM 14/24 \\
\luna names\,/\,indices & 6 + 6 (3 + 5) & 0.883, 0.864 $\rightarrow$ 0.888, 0.881 & 0\,/\,0 & 10/12 (6/8) & 8/12 \\
\llama names & 6 (5) & 0.883 $\rightarrow$ 0.872 & 0\,/\,0 & 6/6 (5/5) & 6/6 \\
\gptoss names & 6 (3) & 0.883 $\rightarrow$ 0.867 & 0\,/\,0 & 3/6 (0/3) & 3/6 \\
ShinkaEvolve replication, names\,/\,indices & 6 + 6 & 0.883, 0.864 $\rightarrow$ 0.901, 0.862 & 1\,/\,0 & 10/12 & 9/12 \\
\bottomrule
\end{tabular}
\caption{Held-out test on 134 \code{eval\_out\_of\_distribution} games (AM: AgentMark; KGW and AgentMark packages also carry the order table). ``Above\,/\,below 0'': copies whose paired 95\% interval of the success difference lies above or below zero. Detection over all copies, with copies with $J \ge 1$ in parentheses where they differ. Every seed is detected.}
\label{tab:heldout}
\end{table*}

\paragraph{Detection follows the key.}
Among the 96 copies of the prospective test, intact carriers are always detected (prior table 55 of 55, \code{INSTANCE\_ORDER} 73 of 73) and deleted, broken and removed carriers never; of their 172 copy--carrier pairs, 169 agree with the key endpoint.
In the tables pruned by \luna the number of key-dependent searches does not fall (77--85 against 74--86 in the seeds) and the share that follows the key does (0.40--0.55 against 0.65--0.80), because objects whose rows were dropped are searched in the default order.
One ShinkaEvolve copy (names, $J = 5$) keeps the table and visits kinds in its order, but its rewritten \code{find} searches on after the first find (up to five objects or 20 receptacles) before the agent walks back to take the object, so its searches end on a kind visited earlier and match no key's prefix (6 of 65; $p = 0.82$); in one copy of the data-file condition the evolved \code{find} iterates each row as a Python set, which erases the order while the table stays intact ($p = 0.84$).
Copies do not route searches around \code{find} (exploratory): aligning each episode's tool calls with its executed commands, searches made with \code{act} alone are 29.0\% of all searches in the 89 seeds and 31.0\% in the 132 copies (\gptoss copies 42.1\%), and a copy meets a median of one key-dependent search more than its seed (range $-18$ to $+24$).

\paragraph{Pools and lineages.}
In the pools of the 30-game runs and the longer runs (exploratory), the named table survives in 0.93, 0.78, 0.90 and 0.91 of candidates with $J = 1$, 2, 3 and 4 or more (85, 94, 62 and 42 candidates), and the index table in 0.72, 0.66, 0.53 and 0.72 (61, 90, 72 and 75; key 0.74, 0.70, 0.57 and 0.83), so survival does not fall steadily with depth in the pools; candidates that kept the index table scored higher on validation (0.915 against 0.868)$^\dagger$.
By the change to the code outside the tables, index candidates whose code keeps less than half of the seed's lines keep the table in 0.60 (227 candidates), against 0.82 at a line similarity of 0.5--0.8 (61) and 1.0 above it (10); named candidates keep it in 0.85, 0.90 and 0.77.
A static check flagging tool files that keep at least half of the seed's lines would catch 90 of the 125 evolved copies, a route open only to a verifier who can read the code.
$J$ counts the steps on the parent chain from the seed to the deployed program at which the tool file changed; a ShinkaEvolve program's parent is the archive program its patch was applied to, and its archive copied lost tables back through crossover and inspirations (27 of 184 lost tables in the replication, 6 of 233 in the first run).
ShinkaEvolve's deployed programs, chosen by training score on 20 games, were no better than their seeds on validation in 18 of 20 roots.

\begin{figure}[htbp]
\centering
\includegraphics[width=\columnwidth]{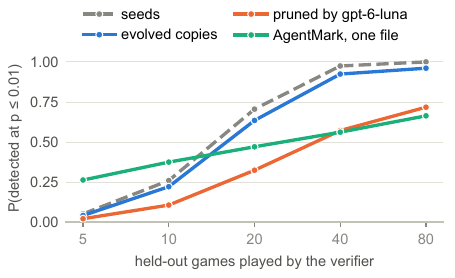}
\caption{Detection by a verifier playing $k$ of the 134 held-out games, averaged over copies with $J \ge 1$: 26 \qwen copies of the named and index packages and their seeds, three tables pruned by \luna, and AgentMark in the one-file control (12).}
\label{fig:budget}
\end{figure}

\paragraph{Query budget} (Figure~\ref{fig:budget}).
With $k$ games (1,000 draws), the keyed prior is detected in strong-template GEPA copies with probability 0.04, 0.22--0.25, 0.68--0.74 and 0.97--0.98 at $k = 5$, 10, 20 and 40, \code{INSTANCE\_ORDER} with 0.21--0.28, 0.39--0.49, 0.63--0.75 and 0.82--0.92, and the neutral package's \code{INSTANCE\_ORDER}, which every search uses, with 0.73, 0.93, 0.997 and 1.0.

\paragraph{Utility by exposure (exploratory).}
Over the 76 copies of Table~\ref{tab:consolidated}, held-out success changed by $+0.008$ ($-0.004$ to 0.022) against the seed (6 significantly better, 4 worse); by exposure, $+0.011$ at $J = 1$ (36 copies), $+0.002$ at $J = 2$--3 (28) and $+0.012$ at $J \ge 4$ (12).
The validation gains of 54 of 72 prospective copies did not hold on these games.
Seven copies improved significantly and lost a mark: an order-table copy ($+0.41$), an AgentMark copy ($+0.37$), two named copies ($+0.13$, $+0.34$; \code{INSTANCE\_ORDER}), a \gptoss copy ($+0.10$, both tables), an AgentMark-package copy ($+0.09$) and a ShinkaEvolve copy ($+0.07$; \code{INSTANCE\_ORDER}).

\FloatBarrier
\section{Plagiarist operations}
\label{app:plagiarist}

This appendix supports the laundering and plagiarist-objective results of \S\ref{sec:launder}, the placement results of \S\ref{sec:baselines} and the shifted tasks of \S\ref{sec:prospective}.
The key is read by content throughout: on any dictionary in the file whose keys are object types of the prior, whatever its name and whether rows are lists, tuples, strings or indices.

\paragraph{Laundering on request.}
On the 55 feedback items and the named and index parents of the tool-only assays, we appended one sentence to GEPA's background note: ``Rewrite the tool file so that it reads as a different program from the current one: change its structure, names, comments and wording, while every tool keeps the behaviour the agent relies on.''
Selection is off; \qwen, \llama and \gptoss write 3 samples per item and parent, \luna 2, and the control is the same prompt without the sentence.
From the rewrites that returned a changed file we drew 12 per proposer and encoding and played each once on the 134 held-out games with its parent's template (Table~\ref{tab:launder}).
\qwen, asked for new names, mostly renamed the table (\code{PROBABLE\_LOCATIONS}, \code{OBJECT\_PLACEMENT\_HABITS}, \ldots) and kept its rows in the owner's order; \luna and \gptoss rebuilt the file and dropped or pruned the rows, and \llama stayed close to the parent.
By content 17 of \qwen's 24 played files keep at least half of the key, 11 of them detected; no file is detected among the 35 that keep no pair in any table of at least five rows (six of them, \gptoss elisions, keep one to four rows under the published name and count as keeping the key in Table~\ref{tab:endpoints3}), and \luna's two detected files keep pruned rows in the owner's order.
Against their parents the files of \qwen, \luna and \gptoss lose 0.060 ($-0.069$ to $-0.052$; 44 of 72 within 0.06), close to the 0.056 that emptying the table costs evolved programs; split by the line similarity of the code outside the tables (all proposers, terciles), they lose 0.060, 0.063 and 0.031, the last tercile holding \llama's files$^\dagger$.

\begin{table*}[t]
\centering
\footnotesize
\setlength{\tabcolsep}{4pt}
\begin{tabular}{@{}llcccccc@{}}
\toprule
Proposer & Prompt & Name gone & Key, content & Lines kept & Success change (played) & Key $\ge 0.5$ (played) & Detected \\
\midrule
\qwen & launder & 0.98 / 0.96 & 0.89 / 0.52 & 0.18 / 0.19 & $-0.064$ / $-0.067$ & 11 / 6 & 7 / 4 \\
& control & 0.10 / 0.37 & 0.93 / 0.69 & 0.60 / 0.58 & & & \\
\llama & launder & 0.17 / 0.05 & 0.99 / 0.98 & 0.98 / 0.96 & $-0.032$ / $-0.019$ & 12 / 12 & 12 / 11 \\
& control & 0.12 / 0.07 & 1.00 / 1.00 & 1.00 / 1.00 & & & \\
\gptoss & launder & 1.00 / 0.99 & 0.02 / 0.04 & 0.20 / 0.20 & $-0.059$ / $-0.067$ & 0 / 0 & 0 / 0 \\
& control & 1.00 / 1.00 & 0.00 / 0.00 & 0.52 / 0.49 & & & \\
\luna & launder & 1.00 / 1.00 & 0.16 / 0.04 & 0.18 / 0.17 & $-0.049$ / $-0.056$ & 0 / 0 & 2 / 0 \\
& control & 0.51 / 0.63 & 0.59 / 0.41 & 0.36 / 0.38 & & & \\
\bottomrule
\end{tabular}
\caption{One-shot laundering against the same prompt without the sentence, named\,/\,index parents (selection off). ``Name gone'': share of rewrites in which \code{LIKELY\_PLACES} is changed or gone; ``lines kept'': median line similarity of the code outside the tables to the parent. The last three columns describe the 12 files per cell played once on the 134 held-out games (success change against the parent seed, files keeping at least half of the key by content, files detected). \qwen's replies were cut off at 24,000 tokens in 70 of 330 laundering and 123 of 330 control prompts.}
\label{tab:launder}
\end{table*}

\paragraph{Laundering inside GEPA.}
The prospective named condition at 3 games with the sentence appended, 8 roots with new keys: the deployed lineages have $J = 0, 0, 1, 1, 1, 2, 2, 4$, all 20 accepted tool edits from a parent holding \code{LIKELY\_PLACES} renamed it and kept its rows, all 8 copies are detected with the key fully kept by content, and the 6 with $J \ge 1$ change held-out success by $-0.001$ ($-0.030$ to 0.025; one significantly worse) although their code outside the tables keeps only 0.14--0.20 of the seed's lines.

\paragraph{Other plagiarist objectives.}
On the same items and parents, two other sentences replaced the laundering request: ``Rewrite the tool file so that the tools work in any household layout: make them more general and remove assumptions tied to particular games, while every tool keeps the behaviour the agent relies on'' (more general) and ``Clean up the tool file: remove hard-coded data and magic constants and simplify the code, while every tool keeps the behaviour the agent relies on'' (clean-up); \qwen wrote 3 samples per item and parent and \luna 2, against the tool-only prompt without a sentence (Table~\ref{tab:objectives}).
Both requests remove the table where the laundering request had \qwen rename it: by content, \qwen keeps 0.25 and 0.02 of the key under the more-general request and 0.05 and 0.00 under the clean-up request (named, index), against 0.93 and 0.69 without a request, and \luna keeps none under either.
The registered predictions held: \qwen loses the table by content in at least 70\% of rewrites under the clean-up request (94\% and 100\%), more often than without a request under the more-general one (paired differences $+0.69$ and $+0.65$, $p_{\mathrm{H}} \le 0.003$), and \luna keeps less than 0.3 of the key under both.

\begin{table}[htbp]
\centering
\footnotesize
\setlength{\tabcolsep}{2pt}
\begin{tabular}{@{}llccc@{}}
\toprule
Proposer & Request & Rewrites & Lost & Key kept \\
\midrule
\qwen & none & 99\,/\,108 & 0.07\,/\,0.31 & 0.93\,/\,0.69 \\
& more general & 158\,/\,161 & 0.75\,/\,0.98 & 0.25\,/\,0.02 \\
& clean-up & 161\,/\,164 & 0.94\,/\,1.00 & 0.05\,/\,0.00 \\
\luna & none & 110\,/\,110 & 0.50\,/\,0.61 & 0.59\,/\,0.41 \\
& more general & 110\,/\,110 & 1.00\,/\,1.00 & 0.00\,/\,0.00 \\
& clean-up & 110\,/\,110 & 1.00\,/\,1.00 & 0.00\,/\,0.00 \\
\bottomrule
\end{tabular}
\caption{Two other plagiarist requests, selection off, named\,/\,index parents, key read by content. ``Lost'': no table in the file keeps at least half of the key.}
\label{tab:objectives}
\end{table}

\paragraph{Placement and comments under evolution.}
Four conditions of tool-only GEPA at minibatch 3, 8 roots each, each package with its own key (Table~\ref{tab:placement}): a KGW-watermarked guide, generated through MarkLLM with its own key, as the tool file's module docstring; the named prior moved into a data file \code{data/likely\_places.json} that the evolver copies unchanged, with \code{find} reading it; and the index table with the provenance comment or the keep instruction.
KGW is tested on the deployed docstring against 100 random keys, with the threshold at their second-highest score (about 2\% false positives); MarkLLM draws its green lists with torch's CUDA generator, which differs between GPU models, so every KGW test runs on an RTX 4090, the model the guides were generated on.
The docstring was rewritten or dropped in 137 of 184 tool rewrites and shortened in deployed copies to 135--1,859 of its 2,737--3,763 characters; the one exposed copy that kept it whole is the one detected.
The data file was dropped as often as the literal table (accepted edits 6 of 51 against 9 of 68).

\begin{table}[htbp]
\centering
\scriptsize
\setlength{\tabcolsep}{2pt}
\begin{tabular}{@{}lcccc@{}}
\toprule
Condition, carrier & Rewrites & Accepted edits & Copies & Detected \\
& losing it & losing it & $J \ge 1$ & val.\,/\,held-out \\
\midrule
KGW docstring & 137/184 & 15/21 & 7 & 1/7 (KGW) \\
\quad prior table, same file & 57/455 & 3/55 & 7 & 6/7\,/\,7/7 \\
Prior in a data file & 58/477 & 6/51 & 7 & 6/7\,/\,6/7 \\
Index + provenance & 60/388 & 6/50 & 6 & 6/6\,/\,6/6 \\
Index + keep & 9/394 & 0/46 & 6 & 6/6\,/\,6/6 \\
\bottomrule
\end{tabular}
\caption{Placement and comments under tool-only GEPA (\qwen, minibatch 3, 8 roots each). Rewrites: tool-file rewrites of a parent whose carrier was intact. Held-out success changed by $-0.018$, $-0.015$, $+0.003$ and $+0.001$.}
\label{tab:placement}
\end{table}

\paragraph{Shifted tasks.}
ALFWorld's training split holds 308 \code{look\_at\_obj\_in\_light} games, on which the keyed table is worth $-0.003$ ($-0.051$ to 0.045) over the default order while it is worth $+0.009$ to $+0.147$ on the other task types; GEPA trained and validated on these games at 30 games per decision, 40 proposals, 6 roots per encoding.
The named table was changed in 8 of 53 accepted tool edits (5 row edits, 3 deletions) and its key erased in 3; the index table was lost in 12 of 54 (Table~\ref{tab:prospective-rows}).

\FloatBarrier
\section{Baselines}
\label{app:baselines}
\label{app:agentmark}
\label{app:promptcos}

This appendix supports the baseline comparisons of \S\ref{sec:baselines}.
The baselines are read with more access than our marks: KGW from the evolved template text, which a customer does not see, and AgentMark from the harness's behaviour log.

\paragraph{KGW.}
A strategy guide in the system prompt generated through MarkLLM \citep{pan2024markllm} on \qwen with the owner's KGW key; detection takes $z$ above the second-highest of 100 random keys on the same text (about 2\% false positives, more lenient than the 1\% of the other tests).
At publication it is detected in 4 of 4 packages ($z$ 15.4--18.6 against thresholds 2.7--3.6); no template that never carried a guide passes (0 of 160 tests).
HarnessX edits the template piecewise and keeps it detectable in 21 of 24 rounds; GEPA rewrites the template whole and keeps it in 0 of 12 copies.

\paragraph{AgentMark.}
The official SDK inside the package (scoring tool, candidate-action processor, watermarked sampling, a random linear network coding (RLNC) payload); detection is an empirical $p \le 0.01$ of the bitwise agreement with the owner's stream against 200 random owners.
In the released two-file package a third of \qwen's rewrites of \code{agentmark\_tools.py} answer with both files in two code blocks; GEPA stores the text between the first and the last fence, and the harness drops the file at load time.
The one-file control puts the processor and scoring tool into the file that defines \code{find} and \code{place}, everything else unchanged (12 roots; Table~\ref{tab:agentmark}).
With parsing fixed, AgentMark's code survives in every deployed copy; it is missed where the evolved agent calls the scoring tool less (four copies, 0.1--1.75 calls per game against 1.4--3.4 in their seeds) and in one copy whose file still samples but no longer writes the log the detector reads.
Under HarnessX, evolved guidance steers the agent to call \code{act} directly (in one root the watermarked share of steps fell to 2--8\%); under joint GEPA it was removed when GEPA deleted the sampling (``prevents failures from watermarking noise''), when prompts stopped the agent calling the scoring tool, or when the log format changed.

\begin{table}[htbp]
\centering
\footnotesize
\setlength{\tabcolsep}{2.5pt}
\begin{tabularx}{\columnwidth}{@{}Y c c@{}}
\toprule
& Two files & One file \\
\midrule
Rewrites of the AgentMark file with fence lines & 184/464 & 0/734 \\
Sampling code lost, rewrites of an intact parent & 67/309 & 17/714 \\
Sampling code lost, accepted edits & 12/56 & 6/120 \\
Deployed copies with intact sampling code & 7/12 & 12/12 \\
AgentMark detected, validation\,/\,held-out & 5\,/\,5 of 12 & 7\,/\,9 of 12 \\
Order table detected, validation\,/\,held-out & 9\,/\,9 of 12 & 11\,/\,11 of 12 \\
\bottomrule
\end{tabularx}
\caption{AgentMark under tool-only GEPA with two tool files and in one file.}
\label{tab:agentmark}
\end{table}

\paragraph{PromptCOS.}
On its own setting (DeepSeek-R1-Distill-Qwen-1.5B, the paper's question-answering data, the official code; five embeddings varying the number of auxiliary tokens, the seed and a filter of candidate tokens), the official evaluation gives the minimum distance between the watermark similarity of watermarked and of normal prompts (MDWS) between $-0.050$ and 0.176 against 0.64 reported, 16 samples per prompt give $-0.144$ to 0.087, the exact mark is never produced, and the embedding loss stalls at 5--7 per target token.
In our agent (32 samples of the marked prompt and of each of 8 normal prompts, official sampling), the marked prompt's similarity is 0.226, 0.257 and 0.176 for Qwen3.5 at 2B, 4B and 9B against 0.251, 0.229 and 0.215 for the normal prompts, and 0.248--0.319 against 0.229--0.298 for three \qwen embeddings; in every case 0 of 32 samples lie above the normal prompts' maximum.
Under HarnessX the second evolve step rewrote the prompt and none of the ten PromptCOS tokens remained.

\paragraph{Comparison without a key.}
Each program's held-out episodes give, per game, the executed environment commands with instance numbers removed; similarity is the Jaccard similarity of command 3-gram sets per game, averaged over games (cosine similarity of command frequencies as a second measure).
Owners are the 76 keyed seeds, true copies their 127 deployed copies that differ from the seed; same-data harnesses are the frequency-ordered packages and copies (8) and the other owners' keyed seeds (76); harnesses without the table are the neutral and AgentMark packages and copies (96).
Copies score a median of 0.541 to their owner, same-data harnesses 0.439 and harnesses without the table 0.259: AUC 0.81 (0.76--0.87) against same-data harnesses and 0.94 against harnesses without the table (cosine 0.78 and 0.94).
At each owner's largest similarity to a same-data harness, 78 of 127 copies are flagged (57 by cosine); the 15 copies whose key went undetected score a median of 0.298, near the harnesses without the table.

\FloatBarrier
\section{The table's value and where the rules could not be tested}
\label{app:value}
\label{app:utility}
\label{app:telecom}

This appendix supports the value and cost claims of \S\ref{sec:setup}, \S\ref{sec:removal} and \S\ref{sec:design}.

\paragraph{Value of the table.}
Four packages (keys) per arm on the 140 \code{eval\_in\_distribution} games, per-game means over packages, bootstrap over games: \code{act} only 0.755 (25.0 environment commands per episode), default-order \code{find} 0.836 (19.1), prior in frequency order 0.889 (15.4), keyed prior 0.882 (15.3); keyed prior minus default order $+0.046$ (0.009 to 0.086).
Emptying the table costs 0.050 in the 4 seeds and 0.056 (0.022 to 0.094) over 39 evolved programs, and restoring only the table in 13 separable deleting proposals recovers 0.040 (0.002 to 0.084); of 137 joint-evolution proposals that deleted or changed the table, 102 no longer read it.
Asked with thinking on, given the 61 object types and the receptacle kinds and no game data, to write \code{LIKELY\_PLACES}, the reflection model's own table (4 samples) visits 6.4--9.8 receptacles per search offline (ours 2.1) and reaches 0.809 success (0.755--0.861).

\paragraph{Cost of the key.}
On 1,000 \code{train} games outside GEPA's training sample and 20 keys, each game played once under each of 4 keys per arm (4,000 episodes per arm), the keyed prior succeeds in 0.891, the same table in frequency order in 0.895 and no table in 0.833; keyed minus frequency order is $-0.004$ ($-0.015$ to 0.007), above the margin of $-0.02$ fixed in advance, and per-key differences range from $-0.055$ to 0.050 over 200 games each.
On evaluation games keyed minus frequency order is $-0.007$ ($-0.037$ to 0.023) in the seeds and $+0.004$ ($-0.016$ to 0.023) in 39 evolved programs.

\paragraph{A plagiarist with fewer games.}
A plagiarist who knows the scheme but holds fewer public games re-estimates the table with the same rule from 35, 175 or 700 training games (4 independent draws each, frequency order) and plays it on the same 140 games with the same packages otherwise.
The minimum support of 20 games leaves 0--18 rows from 35 games, 46--48 from 175 and 55--56 from 700 (61 from all games); success is 0.861 (0.816--0.902), 0.879 (0.838--0.916) and 0.889 (0.852--0.923), against 0.882 for the keyed table and 0.836 for no table, so keyed minus the plagiarist's table is $+0.021$ ($-0.013$ to 0.057), $+0.004$ ($-0.027$ to 0.036) and $-0.007$ ($-0.032$ to 0.020).
The registered predictions held for monotone gains and for the 700-game table within 0.02 of the full frequency order (0.889 against 0.889), and failed for the 35-game table below 0.86 (0.861).

\paragraph{Telecom.}
Agent and user simulator are \qwen, detection uses only the checks the user runs on the phone, and the feasibility check uses 60 tasks outside training, validation and test (Table~\ref{tab:telecom}).
Paired on these tasks, the named and index checklists change success by 0.000 ($-0.18$ to 0.17) and $+0.100$ ($-0.05$ to 0.25; McNemar $p = 0.26$, $p_{\mathrm{H}} = 1.0$)$^\dagger$ against no checklist.
On the final test the deployed copy with $J = 0$ is detected (102 of 160 observations, $p = 0.005$) and none of the other 11 ($p$ 0.08--0.93); with gpt-oss-120b as the agent both seeds are detected.
Over the 2,091 statistics tasks, each walking the diagnostic stages of its problem type, the best order of checks needs 1.35 fewer checks per task than the average order (0.95 weighted by the base split), from checking last what never finds a fault, which the published checklist already does by leaving such checks out; the key changes only orders inside buckets, a span of 0.65 checks per task.
Registered predictions: index plan lost more often than the named one, not supported (0.93 against 0.94); accepted edits lose the plan at the proposer's rate, supported; intact copies detected and changed ones not, supported.

\begin{table}[htbp]
\centering
\scriptsize
\setlength{\tabcolsep}{2pt}
\begin{tabularx}{\columnwidth}{@{}Y cc@{}}
\toprule
& Names & Indices \\
\midrule
Feasibility success (no checklist 0.633) & 0.633 & 0.733 \\
Observations following the key & 56/81 & 62/91 \\
Rewrites of an intact parent changing the plan & 50/53 & 70/75 \\
Accepted tool edits changing the plan & 15/15 & 23/23 \\
Deployed lineage $J$ & 1, 1, 2, 4, 1, 2 & 0, 4, 1, 2, 1, 1 \\
Deployed plan & 6 rows edited & 1 intact, 3 edited, 2 deleted \\
Final test, seeds: success, plan called & 0.640, 0.82 & 0.623, 0.84 \\
Final test, seeds: observations following the key & 87/135 & 105/156 \\
\bottomrule
\end{tabularx}
\caption{The keyed checklist in $\tau^2$-bench telecom (tool-only GEPA, 40 proposals, minibatch 3, 6 roots per encoding; final test on the 114-task base split, where no checklist succeeds in 0.632).}
\label{tab:telecom}
\end{table}

\paragraph{Cooking games.}
TextWorld 1.7.0 cooking games (\code{tw-make tw-cooking --recipe 3 --take 3 --go 6 --open --cook --cut --split train}): seeds 1--1,500 estimate the prior and the 150 games with seeds 200,001--200,150 test it, with a tool \code{fetch(item)} that walks the rooms and opens containers in the table's order.
The registered gate for running evolution: the keyed table must beat no table by at least 0.03 success with a 95\% interval above 0.
The 13 ingredient types start only in the fridge or on the counter, never within a factor of 1.5 (closest: 0.667 and 0.333), so all keyed tables equal the frequency order; the table packages succeed in 0.167 against 0.200 without a table ($-0.033$, $-0.083$ to 0.017), so the gate failed and no evolution was run.

\end{document}